\documentclass[11pt,oneside]{article}
\usepackage{setspace,graphicx,amssymb,amsmath,latexsym,amsfonts,amscd,amsthm,multirow,ctable,mathdots,caption,array,color}
\usepackage{fancyhdr,tabularx,cite,mathrsfs}
\usepackage[headings]{fullpage}
\usepackage{stmaryrd}
\usepackage{rotating}
\usepackage{authblk}
\usepackage{hyperref}
\usepackage{accents}

\usepackage{geometry}

\newcommand{\bea}{\begin{eqnarray}} 
\newcommand{\eea}{\end{eqnarray}} 
\newcommand{\bee}{\begin{eqnarray*}} 
\newcommand{\eee}{\end{eqnarray*}} 
\newcommand{\al}{\begin{align*}} 
\newcommand{\eal}{\end{align*}} 
\newcommand{\be}{\begin{equation}} 
\newcommand{\ee}{\end{equation}} 
 
\newcommand{\bem}{\begin{pmatrix}} 
\newcommand{\eem}{\end{pmatrix}}

\theoremstyle{definition}

\theoremstyle{remark}

\numberwithin{equation}{section}

\usepackage[hang]{footmisc}
\usepackage{lipsum}
\begin{document}

\setstretch{1.4}

\title{
	\vspace{-35pt}
	\textsc{\huge{Mixed Rademacher and BPS black holes}}
}

\author[1]{\small{Francesca Ferrari}\thanks{f.ferrari@uva.nl}}

\author[2]{\small{Valentin Reys}\thanks{valentin.reys@unimib.it}}

\date{}

\affil[1]{Institute for Theoretical Physics Amsterdam and Delta Institute for Theoretical Physics, University of Amsterdam, Science Park 904, 1098 XH Amsterdam, The Netherlands}	
\affil[2]{Dipartimento di Fisica, Universit\`{a} di Milano-Bicocca and  INFN, Sezione di Milano-Bicocca, Piazza della Scienza 3, 20126 Milano, Italy}

\maketitle

\abstract{
\noindent Dyonic 1/4-BPS states in Type IIB string theory compactified on $\mathrm{K}3 \times T^2$ are counted by meromorphic Jacobi forms. The finite parts of these functions, which are mixed mock Jacobi forms, account for the degeneracy of states stable throughout the moduli space of the compactification. In this paper, we obtain an exact asymptotic expansion for their Fourier coefficients, refining the Hardy-Ramanujan-Littlewood circle method to deal with their mixed-mock character. The result is {compared to} a low-energy supergravity computation of the exact entropy of extremal dyonic 1/4-BPS single-centered black holes, obtained by applying supersymmetric localization techniques to the quantum entropy function.
}

\newpage

\tableofcontents
  
\newpage

\section{Introduction}

In striving to describe the microstates of BPS black holes in four-dimensional supergravity theories, valuable insights have been gained from the partition functions of BPS states in the corresponding string theory compactifications. These partition functions can often be related to a topological invariant, whose computation at weak string coupling is able to describe, when going to strong coupling, the exact degeneracies of BPS black holes. It has also been realized that these degeneracies are nicely encoded in the Fourier coefficients of so-called counting functions, which in some cases can be computed exactly. Perhaps the most famous example is that of Type IIB string theory compactified on a six-torus, where the counting function of 1/8-BPS states is a (weak) Jacobi form \cite{Maldacena:1999bp, Shih:2005qf}. Such forms have been extensively studied in analytic number theory and an exact formula for all their Fourier coefficients, the Rademacher expansion\cite{Radem, RadZuc, Niebur1}, is known. This expansion is made possible due to the strong constraints imposed on Jacobi forms stemming from modularity. In the toroidal compactification of Type IIB, the Fourier coefficients of the counting function reproduce, at leading order, the celebrated Bekenstein-Hawking entropy of the corresponding dyonic 1/8-BPS black holes \cite{Shih:2005qf}. But it has also been shown, using supersymmetric localization techniques directly in the low-energy supergravity theory, that all perturbative and non-perturbative corrections to this entropy can be captured and studied analytically \cite{Dabholkar:2010uh, Dabholkar:2011ec, Dabholkar:2014ema}. Such a study revealed an exact match between the string theory and supergravity degeneracies of 1/8-BPS states.\\ 
It is of course interesting to ask whether a similar situation occurs in other types of compactifications. For instance, a more intricate case to examine is the one of Type IIB string theory compactified on $\mathrm{K}3 \times T^2$. The counting function of 1/4-BPS states (dyons) which are stable throughout the compactification's moduli space in this theory \cite{Dijkgraaf:1996it, Shih:2005uc, David:2006yn, Sen:2007, Cheng:2008fc, Gaiotto:2005hc, Dabholkar:2007vk, ChengVerlinde} is related to mixed mock Jacobi forms \cite{Dabholkar:2012nd}. The mixed-mock character of these functions is a consequence of the wall-crossing phenomenon and implies that the 1/4-BPS states counting problem can be translated into the question of recovering the exact Fourier coefficients of certain mixed mock Jacobi forms. \\
In general, mock Jacobi forms can be decomposed into vector-valued mock modular forms. Mock modular forms are characterized by a pair of functions, the form itself and its shadow \cite{Zweger, Zagiermock}. Adding a certain non-holomorphic integral of the shadow to the form restores modularity at the expense of holomorphicity. When the shadow is a cusp form, the Rademacher series can be applied to recover the Fourier coefficients of mock Jacobi forms \cite{Niebur1, Pribit, DunFre, Cheng2011, Whalen}. However, the mock Jacobi forms arising in the counting problem of 1/4-BPS states in Type IIB string theory on $\mathrm{K}3 \times T^2$ have mixed mock components, and their shadows are not cusp forms. Such mixed mock modular forms have been much less studied in the literature, with the notable exceptions of \cite{Bringmann:2010sd, BringMahl}. In this paper, we follow the method introduced by Bringmann and Manschot \cite{Bringmann:2010sd} to obtain an exact asymptotic expansion for the Fourier coefficients of the mixed mock Jacobi forms relevant to the string theory states counting problem.\\ 
Our result can be viewed as an exact formula in inverse powers of the charges for the degeneracy of states making up 1/4-BPS black holes in an $\mathcal{N}=4$ theory of supergravity. Similarly to the 1/8-BPS black holes in $\mathcal{N}=8$ supergravity, a localization computation can also be conducted to determine the exact quantum entropy of such black holes directly in the low-energy effective theory. Progress in this direction has been reported in \cite{Murthy:2015zzy}. It is therefore of interest to compare the macroscopic results obtained using this method to the exact microscopic result derived in this paper. We will explain, however, that some aspects of the supergravity computation are still lacking to conduct a precise comparison on par with the toroidal case. This points to interesting questions regarding localization computations in the context of supergravity which should be examined and answered in order to obtain a complete understanding of the microstates of 1/4-BPS black holes in line with their description as bound states of D-branes.\\ \\
The paper is organized as follows: in section \ref{sec:string-sugra}, we review the matching between the microscopic and macroscopic derivations for the degeneracies of 1/8-BPS states in string theory and supergravity, which relies on a complete understanding of Fourier coefficients of Jacobi forms on one hand, and the calculation of the exact quantum entropy of supersymmetric black holes on the other. We then explain how these concepts generalize to 1/4-BPS states in string theory compactified on $\mathrm{K}3\times T^2$ and $\mathcal{N}=4$ supergravity. In section \ref{sec:Rad}, we introduce the relevant mock Jacobi forms for the string theory counting problem and explain how a generalization of the circle method allows for an exact derivation of their Fourier coefficients. The main result is given in \eqref{eq:mixed-mock-coeffs}. In section \ref{sec:sugra-comp}, we examine the macroscopic computation of the exact entropy of 1/4-BPS black holes and explain which aspects, in our opinion, remain to be understood to obtain a complete match with the microscopic results. We close with some comments in section \ref{sec:conclusion}. Three appendices contain some technical facts of our derivation. Appendix \ref{app:mod-misc} collects relevant aspects of the general theory of (mock) Jacobi forms, appendix \ref{sec:app-asymptotic} examines the asymptotic of our result for the Fourier coefficients of mock Jacobi forms in detail, and appendix \ref{sec:app-Iu} contains explicit formulae related to the macroscopic derivation of the exact entropy of 1/4-BPS black holes.

\section{Microscopic and macroscopic counting of BPS states}
\label{sec:string-sugra}
In this section we summarize and compare the microscopic derivation of the counting functions for 1/8-BPS states in Type IIB string theory compactified on~$T^6$ and 1/4-BPS states in Type IIB string theory compactified on~$\mathrm{K}3\times T^2$. The discussion is followed by a survey of the macroscopic computation of the exact entropy of the corresponding BPS black holes in the low-energy effective supergravity theories. 

\subsection{Type IIB string theory on~$T^6$ and on~$\mathrm{K}3\times T^2$}

Consider Type IIB string theory compactified on a six-torus~$T^6$. Write the torus as~$T^6 = T^4 \times S^1 \times \widetilde{S}^1$ and take the following brane/momentum configuration\cite{Shih:2005qf}: $Q_5$ D5-branes wrapping~$T^4 \times S^1$, $Q_1$ D1-branes wrapping~$S^1$, momentum $n$ along~$S^1$, momentum $\ell$ along~$\widetilde{S}^1$ and one unit of Kaluza-Klein monopole charge on~$\widetilde{S}^1$. \\
The resulting four-dimensional~$\mathcal{N}=8$ string theory has a U-duality group~$E_{7,7}(\mathbb{Z})$ with a unique quartic invariant~$\Delta := Q_e^2\,Q_m^2 - (Q_e \cdot Q_m)^2 = 4mn - \ell^2$ \cite{Cremmer:1979up, Kallosh:1996uy}, where the electric and magnetic charge vectors are given by
\begin{equation}
\label{eq:1/8-BPS-charges}
Q_e^2 = 2n \, , \quad Q_m^2 = 2m := 2\,Q_1Q_5 \, , \quad Q_e\cdot Q_m = \ell.
\end{equation}
In this theory, we are interested in counting the degeneracy of 1/8-BPS states. When the $S^1$ circle is large (compared to the Planck scale) and the $\widetilde{S}^1$ is not yet compactified, the generating function of such states can be obtained from the modified\footnote{The standard elliptic genus for such a theory vanishes due to the presence of fermionic zero-modes.}  elliptic genus of the $(4,4)$ two-dimensional SCFT living on the world-volume of the D1-branes, whose low-energy dynamics is described by a sigma-model with target-space a deformation of the symmetric product $\mathrm{Sym}^m(T^4):= (T^4)^m/S_m$ \cite{Maldacena:1999bp}. This modified elliptic genus is given by
\begin{equation}
\label{eq:gen-funct-EG-T4}
\mathcal{E}^{(m)}_{T^4}(q,\bar{q},y) = \mathrm{Tr}_{\,\mathrm{Sym}^m(T^4)}\Bigl[(-1)^{J_0 - \tilde{J_0}}\,(\tilde{J_0})^2\,q^{\,L_0}\,\bar{q}^{\,\bar{L}_0}\,y^{\,J_0}\Bigr] \, ,
\end{equation}
where the trace is taken over the Ramond-Ramond sector of the Hilbert space of the theory, $J_0$ and $\widetilde{J}_0$ denote the left- and right-moving R-charges, and $L_0$, $\bar{L}_0$ are the Virasoro generators. Through the 4d-5d lift \cite{Gaiotto:2005gf}, the 4d counting function\footnote{The result was determined for $m$ and $n$ coprime, otherwise the existence of bound states of D-branes at threshold obscures the interpretation, see \cite{Maldacena:1999bp}.} of 1/8-BPS states in a compactification with $m=Q_1Q_5$ D-branes is derived directly from the definition of the modified elliptic genus \cite{ Shih:2005qf}.\\
Restricting to the $m=1$ case, the counting function of 1/8-BPS states turns out to be the weak Jacobi form of weight~$-2$ and index~$1$
\begin{equation}
\label{eq:phi-21}
\varphi_{-2,1}(\tau,z) := \frac{\vartheta_1(\tau,z)^2}{\eta(\tau)^6} = \sum_{n\geq0,\ell\in\mathbb{Z}}\,C(4n-\ell^2)\,q^n\,y^\ell \, .
\end{equation}
Above, we use the standard notation~$q := e^{2\pi i \tau}$ and~$y := e^{2\pi i z}$ and the functions $\vartheta_1(\tau,z)$ and $\eta(\tau)$ are defined in Appendix \ref{app:mod-misc}, where we also collect standard facts about Jacobi forms. Notice that the $\bar{q}$ dependence of the modified elliptic genus \eqref{eq:gen-funct-EG-T4} has dropped out since only the right-moving ground states contribute \cite{Maldacena:1999bp}. Taking into account the 4d-5d lift, the degeneracies of 1/8-BPS states are related to the Fourier coefficients of $\varphi_{-2,1}$ by
\begin{equation}
\label{eq:1/8-BPS-degen}
d_{1/8}(\Delta) = (-1)^{\Delta + 1}\,C(\Delta) \, .
\end{equation}
The fact that the degeneracies only depend on the invariant $\Delta = 4n-\ell^2$ is consistent with the $\mathrm{E}_{7,7}(\mathbb{R})$ symmetry of the low-energy supergravity theory\footnote{From a number-theoretic perspective, this is due to the elliptic property of a (weak) Jacobi form which implies that for even weight and prime power index, the Fourier coefficients only depend on $\Delta$.}. In conclusion, the counting of 1/8-BPS states can be obtained in closed form once the Fourier coefficients of the weak Jacobi form \eqref{eq:phi-21} are known.\\ \\
If instead of the Type IIB system on $T^6$ we consider Type IIB compactified on~$\mathrm{K}3\times T^2$, we obtain an~$\mathcal{N}=4$ string theory in four dimensions. Writing~$T^2 = S^1\times \widetilde{S}^1$, we focus on the following duality frame \cite{Shih:2005uc}: $Q_5$ D5-branes wrapping~$\mathrm{K}3\times S^1$, $Q_1$ D1-branes wrapping~$S^1$, momentum~$n$ along~$S^1$, momentum~$\ell$ along~$\widetilde{S}^1$ and one unit of KK monopole charge on~$\widetilde{S}^1$.\\
In this theory, we are interested in the degeneracies of 1/4-BPS states when the $S^1$ circle is large. The counting function for such states in the 5d theory is obtained from the elliptic genus of the (4,4) two-dimensional SCFT with target-space $\mathrm{Sym}^{m+1}(\mathrm{K}3)$ living on the D1-branes world-volume. The latter can be recovered via the multiplicative lift of the elliptic genus for a similar sigma-model with a single $\mathrm{K}3$ as target-space \cite{Dijkgraaf:1996it, Dijkgraaf:1996xw}, that is to say the elliptic genus of the $\mathrm{K}3$ surface
\be
\mathcal{E}_{\mathrm{K}3}(q,\bar{q},y) = \mathrm{Tr}_{\,\mathrm{K}3}\Bigl[(-1)^{J_0 - \tilde{J_0}}\,\,q^{\,L_0}\,\bar{q}^{\,\bar{L}_0}\,y^{\,J_0}\Bigr] = 2\,\varphi_{0,1}(\tau,z)\, ,
\ee
with $\varphi_{0,1}$ defined in \eqref{eq:app-phi0.1}. Moreover, the 4d-5d lift introduces an additional factor corresponding to the center-of-mass degrees of freedom of the bound state of D-branes, which is expressed in terms of the inverse of the standard Jacobi form $\varphi_{10,1}$ \eqref{eq:app-phi10.1} \cite{Shih:2005uc, David:2006yn}. The complete result can be written in terms of an automorphic form, and is given by
\begin{equation}
Z_{\textrm{1/4-BPS}}(\tau,z,\sigma) = \frac{1}{\Phi_{10}(\tau,z,\sigma)} \, 
\end{equation}
the inverse of the Igusa cusp form. This meromorphic Siegel modular form\footnote{The reader is referred to \cite{Geer} for a detailed review on Siegel modular forms.} encodes the degeneracies of 1/4-BPS states. However, to extract the relevant information for the dyonic degeneracy a precise procedure has to be implemented~\cite{Dabholkar:2012nd}.
First, we Fourier expand the counting function in the~$\sigma$ variable\footnote{Recall that the grading $(n,m,\ell)$ corresponds to the $T$-duality invariant integers associated to the chemical potentials $(\tau,\sigma,z)$.}:
\begin{equation}
\frac{1}{\Phi_{10}(\tau,z,\sigma)} = \sum_{m\geq-1}\psi_m(\tau,z)\,e^{2\pi i m \sigma} \, .
\end{equation}
The functions~$\psi_m$ are meromorphic Jacobi forms of weight $-10$ and index~$m$. This is one of the crucial differences with the toroidal case: the $\psi_m$ are \emph{meromorphic} functions with a double pole at~$z=0$ (together with all the points differing by a translation of the period lattice) while $\varphi_{-2,1}$ is a weak Jacobi form holomorphic in $z$. Notice that the meromorphicity of the counting function is an intrinsic characteristic of the 4d theory, and is absent from the 5d picture.\\
The meromorphic Jacobi forms $\psi_m$ can be canonically split into two terms, according to~\cite{Zweger, Dabholkar:2012nd}
\begin{equation}
\label{mero-split}
\psi_m(\tau,z) = \psi_m^F(\tau,z) + \psi_m^P(\tau,z) \, , 
\end{equation}
where~$\psi_m^F$ are mock Jacobi forms\footnote{$\psi_m^F$ are mixed mock Jacobi forms, however throughout the text we will simply write mock Jacobi forms to refer to forms with generic shadow and explicitly use the attribute mixed when the distinction is necessary.}, a concept we recall in Appendix \ref{sec:app-Jac}, and $\psi_m^P$ are meromorphic mock Jacobi forms, which account for the polar part of $\psi_m$
\begin{equation}
\label{eq:polar-piece}
\psi_m^P(\tau,z) = \frac{p_{24}(m+1)}{\Delta(\tau)}\sum_{s\in\mathbb{Z}}\frac{q^{ms^2+s}y^{2ms+1}}{(1-y q^s)^2} \, .
\end{equation}
Above $\Delta(\tau)=\eta(\tau)^{24}$ is the discriminant function, $p_{24}(m+1)$ denotes the coefficient in $\Delta(\tau)^{-1}$ of $q^m$ and the second term is the Appell-Lerch sum of weight 2 and index $m$.
Physically, the splitting~\eqref{mero-split} elucidates the interpretation of the wall-crossing phenomenon\cite{ChengVerlinde, Denef:2007vg, Sen:2007,Cheng:2008fc, Gaiotto:2005hc, Dabholkar:2007vk, David:2006ji,David:2006yn} in the low-energy supergravity theory. Contrary to the toroidal case, the $\mathrm{K}3 \times T^2$ compactification allows for multi-centered configurations of bound states of black holes \cite{Denef:2000nb, Dabholkar:2009dq}. These configurations are stable in certain regions of the moduli space of the supergravity theory, and unstable in others. The splitting~\eqref{mero-split} takes this into account: the piece~$\psi_m^F$ contains the degeneracies of the single-centered immortal gravitational configurations, which are stable throughout the moduli space and hence devoid of poles, while~$\psi_m^P$ encodes the physics related to multi-centered gravitational configurations decaying or appearing when tuning the moduli. The ability of the latter to capture all the walls of marginal stability is built-in by the averaging over $s$, the spectral flow variable, in \eqref{eq:polar-piece}.\\
In this paper, we will focus on the single-centered contributions\footnote{The multi-centered configurations have not been analyzed, as of yet, in the supergravity theory in as much detail as what one can do for the single-centered ones.}, whose degeneracy is captured by the Fourier coefficients of the mock Jacobi forms~$\psi_m^F$.\\ \\
Before analyzing these Fourier coefficients in detail, we first review the derivation of the counting functions for $1/8$-BPS states in the $T^6$ theory and for $1/4$-BPS states in the $\mathrm{K}3 \times T^2$ theory from the low-energy, effective theory perspective. Indeed, it is in principle possible to recover the degeneracies of BPS states in string theory by computing the entropy of BPS black holes in supergravity. In the large charge (thermodynamic) limit, this has been thoroughly investigated, starting with the celebrated result of \cite{Strominger:1996sh}. However, it is also well-known that the Bekenstein-Hawking entropy of BPS black holes receives quantum corrections which should be examined carefully. Thanks in part to supersymmetry, such corrections can be computed exactly by means of the quantum entropy function \cite{Sen:2008vm}. This formalism goes beyond the leading entropy contribution by re-summing all quantum corrections, and in certain cases turns out to be able to reproduce the exact degeneracies of BPS states described by the above modular objects. 

\subsection{The quantum entropy function}
\label{sec:QEF}

Sen motivated a definition of the full quantum-corrected entropy of extremal dyonic black holes in theories of supergravity on the basis of the $AdS_2/\mathrm{CFT}_1$ correspondence \cite{Sen:2008vm}, where the $AdS_2$ factor is a universal factor in the near-horizon region of extremal black holes in any dimension. This quantum entropy of course receives its main contribution from the classical Bekenstein-Hawking entropy, but also encodes the corrections to the area-law coming from higher derivative corrections and quantum fluctuations of the supergravity fields in the near-horizon region. As such, this quantity has a chance of completely reproducing the string-theoretic result for the degeneracy of BPS states which correspond to BPS black holes in the low-energy effective theory.\\ 
According to Sen's definition, the exact entropy of such black holes is formally defined as a path-integral (the expectation value of a Wilson line) on the near-horizon Euclidean $AdS_2$ region,
\begin{equation}
\label{eq:QEF}
\exp[\mathcal{S}_{\mathrm{macro}}](q,p) := W(q,p) = \left\langle\exp\bigl[-\mathrm{i}\,q_I\,\oint\,\mathrm{d}\tau\,A_\tau^I\bigr]\right\rangle_{AdS_2}^\mathrm{finite} \, ,
\end{equation}
where $\tau$ is the (Euclidean) time direction, $A_\mu^I$ are the gauge fields of the vector multiplets under which the black hole is charged with electric charges $q_I$ and magnetic charges $p^I$, and the superscript `finite' denotes a regularization scheme to take care of the infinite volume of $AdS_2$ \cite{Sen:2008vm}. According to the expectations borne out of the attractor mechanism \cite{Ferrara:1995ih}, this exact entropy only depends on the charges of the black hole.\\
For 1/8-BPS black holes in four-dimensional $\mathcal{N}=8$ supergravity with charges corresponding to \eqref{eq:1/8-BPS-charges}, the quantum entropy function $W(q,p)$ is expected to match the microscopic result \eqref{eq:1/8-BPS-degen}, and similarly for 1/4-BPS black holes in 4d, $\mathcal{N}=4$ supergravity. In order to be able to explicitly evaluate the quantum entropy function for these black holes, one regards them as 1/2-BPS solutions of a truncated~$\mathcal{N}=2$ supergravity theory \cite{Shih:2005he}. Their near-horizon region is then full-BPS and is an attractor $AdS_2 \times S^2$ geometry \cite{LopesCardoso:2000qm} with an enhanced~$SL(2) \times SU(2)$ bosonic symmetry. In $\mathcal{N}=2$ supergravity, one can also make use of the superconformal formalism \cite{deWit:1979dzm, deWit:1980lyi} and work with an off-shell formulation of the theory, which turns out to be extremely convenient to apply supersymmetric localization techniques to compute the path-integral \eqref{eq:QEF} \cite{Dabholkar:2010uh, Dabholkar:2011ec}.\\ \\
A large class of~$\mathcal{N}=2$ superconformal gravity actions (F-terms) are entirely specified by a holomorphic function~$F(X^I,\widehat{A})$ called the prepotential, which is a homogenous function of degree 2. Here,~$X^I$ are the scalar fields sitting in the vector multiplets, and~$\widehat{A}$ is the lowest component of the square of the Weyl multiplet (which contains the graviton),~$\widehat{A} = (T_{\mu\nu}^-)^2$. Another class of actions (D-terms) have been showed not to contribute to the quantum entropy function in~\cite{Murthy:2013xpa}, so one can safely focus on the F-terms. For such actions, it was shown in~\cite{Dabholkar:2010uh, Gupta:2012cy} using supersymmetric localization that the leading contribution to $W(q,p)$, denoted by a hat in the formula below and corresponding to the leading saddle-point of the black hole partition function where the Weyl multiplet is at the attractor value $AdS_2 \times S^2$, takes the form
\begin{equation}
\label{eq:QEF-loc}
\widehat{W}(q,p) = \int_{\mathcal{M}_Q} \prod_{I = 0}^{n_v} \; [\mathrm{d}\phi^I] \, \exp\Bigl[-\pi\,q_I \phi^I + 4\pi\,\textnormal{Im} F\bigl(\frac{\phi^I + i p^I}{2}\bigr)\Bigr] Z_{\textnormal{1-loop}}(\phi^I) \, , 
\end{equation}
where
\begin{itemize}
\item $\mathcal{M}_Q$ is the localizing manifold, which is specified by all field configurations satisfying~$Q\Psi = 0$ for all fermions~$\Psi$ in the theory and for the supercharge~$Q$ with the algebra~$Q^2 = L_0 - J_0$, with~$L_0$ the Cartan generator of~$SL(2)$ and~$J_0$ the Cartan generator of~$SU(2)$. 
\item $\phi^I$ are the coordinates on~$\mathcal{M}_Q$, parameterizing the fluctuations of the supergravity fields in the near-horizon $AdS_2$ region,
\item $[\mathrm{d}\phi^I]$ is a measure\footnote{We believe that this measure is subtle and not yet very well understood. We will comment on this below. For an attempt at deriving this measure, see \cite{Gomes:2015xcf}.} taking into account the curvature of~$\mathcal{M}_Q$,
\item $Z_{\textnormal{1-loop}}$ is a one-loop determinant factor arising from quadratic field fluctuations orthogonal to~$\mathcal{M}_Q$ and which depends on the prepotential and the field content of the theory \cite{Murthy:2015yfa, Gupta:2015gga},
\item the prepotential is evaluated at the attractor value~$\widehat{A} = -64$ \cite{LopesCardoso:2000qm}.
\end{itemize}
We stress that this localized form of the quantum entropy function only depends on the prepotential $F(X^I,\widehat{A})$ and the field content (one Weyl multiplet, $n_v$ vector multiplets and $n_h$ hypermultiplets) of the truncated $\mathcal{N}=2$ theory, as well as on the measure $[\mathrm{d}\phi^I]$. This is the only data which must be specified in order to obtain the exact quantum entropy of the 1/8-BPS and 1/4-BPS black holes considered above, and is an upshot of using the superconformal off-shell formalism.\\ \\
For the toroidal compactification outlined at the beginning of this section, the field content is that of $n_v + 1 = 8$ vector multiplets (including the conformal compensator) and $n_h = 0$ hypermultiplets, and the prepotential is given by (see e.g. \cite{Shih:2005he})
\begin{equation}
\label{eq:prepot-T6}
F(X^I) = -\frac{X^1\,X^a\,C_{ab}\,X^b}{X^0} \, , \quad a,b = 2\ldots 7 \, ,
\end{equation}
where $C_{ab}$ is the intersection matrix of the six 2-cycles on $T^4$. Note the absence of terms proportional to the $\widehat{A}$ field, meaning that the supergravity action only contains two-derivative terms in this case. Because of this, the measure $[\mathrm{d}\phi^I]$ can be obtained straightforwardly from the measure of the scalar fields' target-space \cite{Dabholkar:2011ec, Murthy:2013xpa}.\\ \\
For the $\mathrm{K}3 \times T^2$ case, the field content of the truncated $\mathcal{N}=2$ theory is that of $n_v + 1 = 24$ vector multiplets (including the conformal compensator) and $n_h = 0$ hypermultiplets, and the prepotential is given by the exact expression \cite{Harvey:1996ir,Shih:2005he}
\begin{equation}
\label{eq:prepot-K3}
F(X^I,\widehat{A}) = -\frac{X^1\,X^a\,C_{ab}\,X^b}{X^0} - \frac{\widehat{A}}{128\,i\pi}\,\log\eta^{24}\Bigl(\frac{X^1}{X^0}\Bigr) \, , \quad a,b = 2\ldots 23  \, ,
\end{equation}
where $C_{ab}$ is the intersection matrix of the 2-cycles on $\mathrm{K}3$ and $\eta(\tau)$ is the Dedekind eta function \eqref{eq:app-eta}. Observe that in this case, there are brane instanton corrections proportional to the chiral background field $\widehat{A}$ as compared to \eqref{eq:prepot-T6}. The measure on the localizing manifold is also more subtle, and we will comment on this in section \ref{sec:sugra-comp}.\\ \\
In \cite{Dabholkar:2010uh, Dabholkar:2011ec}, the authors focused on the quantum entropy function of a 1/2-BPS black hole in $\mathcal{N}=2$ superconformal gravity with electric charges $q^2 = 2n$ and magnetic charge $p^2 = 2$ describing 1/8-BPS states in the $T^6$ compactification described above (with $m=1$). They showed that (\ref{eq:QEF-loc}) takes the form
\begin{equation}
\widehat{W}(q,p) = 2\pi\,\Bigl(\frac{1}{4n-\ell^2}\Bigr)^{7/4}\,I_{7/2}\Bigl(\pi\sqrt{4n-\ell^2}\Bigr) \, ,
\end{equation}
where $I_\rho(z)$ is the I-Bessel function of weight $\rho$ \eqref{eq:app-Bessel} and $\ell = q\cdot p$. Moreover, sub-leading saddle-points in the quantum entropy function corresponding to orbifolded sectors of the near-horizon region $AdS_2 \times S^2 / \mathbb{Z}_k$ with $k>1$ were also considered in \cite{Dabholkar:2014ema}. These saddle-points encode exponentially suppressed contributions to $W(q,p)$ and the full answer thus takes the form of a sum over geometries reminiscent of the black hole Farey tail \cite{Dijkgraaf:2000fq, ManscMoore, ChengMB}. In fact, this sum turns out to be precisely the low-energy manifestation of the so-called Rademacher series, which is a powerful tool to reconstruct the Fourier coefficients of modular objects. As reviewed in Appendix \ref{app:mod-misc}, the sole knowledge of the polar terms (terms with a strictly negative $q$-power) in the Fourier expansion of a (weak) Jacobi form, together with its modular properties, is enough to obtain an exact formula for all the Fourier coefficients. The expansion parameter corresponds to the orbifold order $k$ in supergravity. In the $T^6$ case, there is a precise match between the coefficients of the counting function $\varphi_{-2,1}$ \eqref{eq:phi-21} obtained using the Rademacher expansion and the total quantum entropy function $W(q,p)$, at each order in the $k$ parameter \cite{Dabholkar:2011ec, Dabholkar:2014ema}. \\
In fact, one can take the point of view that the precise structure provided by the Rademacher expansion offers a steady guide to applying localization techniques to the quantum entropy function. A priori, such a computation is not straightforward and numerous aspects of the supergravity theory under consideration need to be examined in details before one can trust the results. Nevertheless, its ability to completely reproduce the Fourier coefficients of the Jacobi form $\varphi_{-2,1}$ (and thus all perturbative and non-perturbative corrections to the Bekenstein-Hawking entropy) in the toroidal case considerably strengthens the validity of the approach a posteriori.\\ \\
In the less supersymmetric $\mathrm{K}3 \times T^2$ compactification, the computations become more troublesome on both sides (microscopic and macroscopic). Despite the fact that the macroscopic entropy formula \eqref{eq:QEF-loc} can be applied just as in the toroidal case, one has to deal with the instanton contributions to the prepotential \eqref{eq:prepot-K3}. For the leading saddle-point to the quantum entropy function (un-orbifolded near-horizon geometry), this was examined in \cite{Murthy:2015zzy}. One should also examine potential other geometries corresponding to sub-leading saddle-points and understand what is the effect of summing over these in the full $W(q,p)$. Investigations in this direction have been conducted in \cite{Banerjee:2008, Murthy:2009dq}, where it was shown that orbifolded near-horizon geometries also give exponentially suppressed contributions to the entropy in this case. We will come back to the supergravity calculation in section \ref{sec:sugra-comp} and explain how the results obtained in this paper can be used as a guiding principle in a way similar to the invaluable role played by the Rademacher expansion in the toroidal case. \\
On the microscopic side, the counting functions for immortal black holes $\psi_m^F$ are not Jacobi forms but mock Jacobi forms. As we will see in the next section, this means that the Rademacher formula needs to be generalized in order to obtain a convergent asymptotic expansion for their Fourier coefficients.

\section{Rademacher series}
\label{sec:Rad}

In this section, we review the procedure leading to a Rademacher series for (mock) modular forms via the circle method and extend the method to the case of mixed mock modular forms, along the lines of \cite{Bringmann:2010sd}. In the second part, a description of the mock modular objects of interest to us will be followed by the derivation of the Rademacher expansion for their Fourier coefficients.

\subsection{The circle method}
\label{sec:circle-method}

Several techniques are available to compute the asymptotic growth of the Fourier coefficients of modular objects. Among these is the well-known method of steepest descent, which provides an estimate of the asymptotic growth via a saddle-point approximation. In this section we introduce another powerful technique: the Hardy-Ramanujan-Littlewood circle method. The strength of this technique lies in the fact that it exactly reconstructs the Fourier coefficients of the modular object, going beyond the asymptotic growth estimation.\\
The circle method was discovered by Hardy and Ramanujan\cite{HardRam} in the study of the generating function of unrestricted partitions, $\mathcal{P}(q)=1+ \sum_{n>0} \alpha_j(n)q^n$, whose Fourier coefficients are defined by Cauchy's theorem as
\be
\alpha(n) = \frac{1}{2\pi i} \oint_{\gamma} \frac{\mathcal{P}(q)}{q^{n+1}}\,\mathrm{d}q \,,
\ee
where the contour $\gamma$ encircles the origin of the $q$-plane and lies inside the unit circle. For a complete review of the circle method the reader is referred to \cite{RademFourier}. $\mathcal{P}(q)$ is strictly related to the Dedekind eta function $\eta(\tau)$ by the formula
\be
\mathcal{P}(q)= q^{1/24}\eta(\tau)^{-1} =\prod\limits_{n=1}^{\infty} (1-q^n)^{-1}, \quad q=e^{2\pi i \tau}.
\ee
In order to proceed we introduce some notation which will be important later. \\ \\
\emph{\bf Definition}: A (vector-valued) modular form of weight $w \in \frac{1}{2}\mathbb{Z}$ and multiplier system $\rho$ with respect to the modular group $SL(2,\mathbb{Z})$ is a holomorphic function $\vec{f}: \mathbb{H}\rightarrow \mathbb{C}^d$ satisfying the functional equation 
\begin{equation}
\vec{f}(\gamma\tau) = j_w(\gamma,\tau) \rho(\gamma).{\vec{f}}(\tau) \; ,
\end{equation}
where $\gamma\tau$ denotes the usual action of the modular group on $\mathbb{H}$, the dot stands for matrix multiplication and $j_w(\gamma,\tau)=(c\tau+d)^w$ is the automorphic factor, where the principal branch of the logarithm has been chosen. \\
By assumption, the multiplier system $ \rho(\gamma)$ is a map $\rho: SL(2,\mathbb{Z}) \rightarrow U(d)$, such that $\rho(T)$ is a diagonal matrix whose entries are roots of unity\footnote{Here $T$ denotes the upper triangular matrix defined in \eqref{eq:sl2zgen}.}. 
The transitivity of the action of the modular group then follows from the consistency condition
\begin{equation}
j_w(\alpha\beta,\tau)\,\rho(\alpha\beta)= j_w (\alpha,\beta\tau)\,j_w (\beta,\tau)\,\rho(\alpha)\,\rho(\beta) \quad \mathrm{for} \;\; \alpha,\beta \in SL(2,\mathbb{Z}) \, .
\end{equation}
If the Fourier expansion of each component of the vector-valued modular form starts with a positive power of $q$, $\vec{f}(\tau)$ is called a cusp form. 
Clearly, scalar-valued modular forms are vector-valued modular forms with a single component. The Dedekind eta function, introduced above, is a (scalar-valued) modular form of weight $1/2$ under $SL(2,\mathbb{Z})$ with multiplier system satisfying $\rho_{\eta}^{24}=1$, see \eqref{eq:app-eta-mult}. \\ \\
Hardy and Ramanujan noticed that the Fourier coefficients $\alpha(n)$ could be reconstructed following the pole structure of the function $\mathcal{P}(q)$. In other words, the poles of this function indicate the path of integration to follow in order to have the biggest contribution to the integral along a relatively small portion of the path. In the case at hand, the heaviest contribution is given by the pole at $q=1$, then all the different roots of unity appear with decreasing weight for increasing denominator. \\
This led to the decomposition of the circle $\gamma$ into a sum of Farey arcs, 
\be
\oint_{\gamma} \qquad \longrightarrow \quad \sum_{\substack{0\leq h<k\leq N \\ (h,k)=1}} \int_{\xi_{h,k}}
\ee
Here $\xi_{h,k}$ denotes the Farey arc centered in $h/k$ and bounded by $\frac{h+h_1}{k+k_1}$ and $\frac{h+h_2}{k+k_2}$, with $\frac{h_1}{k_1}$ and $\frac{h_2}{k_2}$ indicating the preceding and the consecutive Farey fractions of $h/k$ respectively. 
Indeed, the Farey series of order $N$, $\mathcal{F}_N$ \cite{Fareybook}, is constituted by irreducible fractions $h/k$ in ascending order such that $0\le h<k \le N$, and $(h,k)=1$, i.e. $h$ and $k$ are coprime. \\
This strategy was refined by Hardy and Littlewood \cite{HardyLittl} and adopted by Rademacher to derive an exact expression for the Fourier coefficients of the $J$-function \cite{Radem} and, together with Zuckerman, to prove the form of the Fourier coefficients of certain modular forms with negative weights \cite{RadZuc}. 
Following their discussion, the modular variable $q$ is substituted by $e^{-\frac{2\pi}{N^2} +2\pi i \phi+ 2\pi i\frac{h}{k}}$, such that in the limit $N\rightarrow \infty$ the circle $\gamma$ tends to the unit circle. After having transformed the contour of integration into a sum of Farey arcs and implemented the above substitution, the integral takes the form
\begin{equation}
\label{eqn:circ3}
\alpha(n) = \sum_{\substack{0\leq h<k\leq N \\ (h,k)=1}} e^{-2i\pi n \frac{h}{k}} \int_{-\vartheta'_{h,k}}^{\vartheta''_{h,k}}\, {f}\Bigl(e^{-\frac{2\pi}{N^2} + 2i\pi\frac{h}{k} + 2i\pi\phi}\Bigr)e^{\frac{2\pi n}{N^2} - 2i\pi n\phi}\,\mathrm{d}\phi \, ,  
\end{equation}
where $\vartheta'_{h,k}$ and $\vartheta''_{h,k}$ are the mediants of the Farey arc after the change of variables. The function $f$ represents the $J$-function or a more general modular form; an explanation regarding the type of object one can consider will follow shortly. \\
Consider the variable $z = \frac{k}{N^2} - i k \phi $ and denote an element of $SL(2,\mathbb{Z})$ as
\be
\bem
a & b \\
c & d
\eem= \bem
h' & -\frac{hh'+1}{k} \\
k & -h
\eem \qquad hh' \equiv -1 \, (\text{mod}\, k)
\ee
such that $\tau = \frac{1}{k}(h+iz)$ and $\gamma \tau= \frac{1}{k}(h'+\frac{i}{z})$. Note that $\, z\in \mathbb{C}, \, Re(z)>0 \, $. Furthermore, $k$ may be restricted to take positive values, thanks to the symmetry under $-\mathbb{I}$ of ${f}$. \\ 
Since we focus on the behavior of the function close to the roots of unity, performing a modular transformation on $f$ allows one to obtain an estimate of the function which, in the neighborhoods of roots of unity, is dominated by the polar $q$-terms in the expansion.
Further refinement of the estimate invokes the limit for $N\rightarrow \infty$, and leads to an expression in terms of Kloosterman sums and Bessel functions. For the $J$-function, the Rademacher series is given by the following expression
\be
J(\tau)= q^{-1}+ c_0 + \sum\limits_{n=1}^{\infty} \frac{2\pi}{\sqrt{n}}\sum\limits_{k=1}^{\infty}\frac{Kl(n,-1;k)}{k}I_1\Bigl(\frac{4\pi \sqrt{n}}{k}\Bigr) \, q^n
\ee
where $Kl(n,-1;k)= \sum\limits_{\substack{h\, \text{mod} \,k \\ (h,k)=1}} e^{-\frac{2\pi i}{k}(nh+h')}$ is the so-called Kloosterman sum and $I_1$ denotes the I-Bessel function defined in \eqref{eq:app-Bessel}. Although the constant term $c_0$ might need more care for the proof of convergence, it can be recovered by analyzing the behavior of the Kloosterman sum at $n = 0$\cite{Radem}. \\
The method was later developed for modular forms of various weights, modular groups and multiplier systems in several works like \cite{Knopp1, Knopp2, KnoMas, Niebur1, Pribit, DunFre, Cheng2011, Whalen}, to quote just a few\footnote{In\cite{Cheng2012} a collection of references together with the history of the Rademacher sum from the perspective of Poincar\'{e} sums is presented.}. Progress in the context of harmonic Maass forms has also been achieved in \cite{{BO06},{BO08},{BO10}}.\\
Remarkably, Niebur showed that the functions\footnote{Niebur coined these functions ``{automorphic integrals}".}  whose Fourier coefficients are defined by the Rademacher series satisfy a particular functional equation\cite{Niebur1}. The latter coincides with the modular transformation of a mock modular form with a cusp form as shadow, which we now turn to.\\ \\
Mock modular forms were first investigated by Ramanujan as $q$-hypergeometric series in\cite{Raman}. After the appearance of these objects both in mathematics and physics, they have been interpreted by Zwegers in\cite{Zweger} as mock modular forms, and in particular as mock theta functions in the case of Ramanujan's examples. \\ 
The definition of mock modular forms extends the one of modular forms in the following way: mock modular forms have the peculiarity of transforming as modular forms only after the addition of a non-holomorphic term. \\ \\
\emph{\bf Definition}: A vector-valued mock modular form of weight $w$ and multiplier system $\rho$ with respect to $SL(2,\mathbb{Z})$ is a holomorphic function $\vec{h}(\tau)$ with at most exponential growth at the infinity cusp and such that there exists a non-holomorphic function, whose components are given by 
\be
\widehat{h}_\ell(\tau)= h_\ell(\tau) + g_\ell^*(\tau), 
\ee
and transform as modular forms of weight $w$ and multiplier system $\rho$ over $SL(2,\mathbb{Z})$. The function $\widehat{h}_\ell(\tau)$ is called the \emph{completion} of the mock modular form and it is related to the original function by the addition of the (non-holomorphic) Eichler integral of a modular form $g_\ell(\tau)$, the so-called \emph{shadow}. \\ \\
The Eichler integral of $g_\ell(\tau)$ is defined by\footnote{Note that this representation of the Eichler integral is valid for $w>1$ or for $g_\ell(\tau)$ a cusp form.}
\be
\label{eq:Eichler}
g_\ell^*(\tau)= \Bigl(\frac{i}{2\pi}\Bigr)^{w-1}\int_{-\bar{\tau}}^{\,i\infty}\,(z+\tau)^{-w}\,\overline{g_\ell(-\bar{z})}\,\mathrm{d}z,
\ee
where $g_\ell(\tau)$ is a component of a vector-valued holomorphic modular form of weight $2-w$ and multiplier system conjugate to the one of $h_\ell(\tau)$. Clearly, a modular form is simply a special case of a mock modular form with trivial shadow.\\
Although a Rademacher-type expression already exists for vector-valued mock modular forms with a cusp form as shadow, more general modular objects such as mixed mock modular forms do not fall in this category and our discussion must be adapted consequently. 
Weakly holomorphic mixed mock modular forms lie in the tensor product of modular forms and weakly holomorphic mock modular forms.
Indeed, a \emph{mixed mock modular form}, $h_\ell(\tau)$, is a mock modular form whose completion transforms as a modular form of weight $w$ and takes the form
\be
\widehat{h}_\ell(\tau)= h_\ell(\tau)+ r(\tau)\,g_\ell^*(\tau) 
\ee 
where $r(\tau)$ is a holomorphic modular form of weight $s$ and $g_\ell(\tau)$ is a holomorphic modular form of weight $2-w+s$. \\
Next we will deal with the mixed mock modular forms appearing in the counting problem of immortal dyons in string theory on $\mathrm{K}3\times T^2$. 

\subsection{Mixed Rademacher series}
\label{sec:mixed-Rad}

The aim of the following analysis is to derive an analytic expression for the Fourier coefficients of the mixed mock modular forms arising in the theta-decomposition of the mock Jacobi forms $\psi_m^F$ for different values of $m$, answering the question raised in section \ref{sec:string-sugra} about the construction of a Rademacher expansion for vector-valued mixed mock modular forms.\\ 
Before applying the circle method to these mixed mock modular forms, we examine the precise structure of the mock Jacobi forms under consideration. A detailed treatment of these functions is provided in \cite{Dabholkar:2012nd}. There it was shown that, after multiplication by the discriminant function $\Delta(\tau)$, the mixed mock Jacobi form $\psi_m^F$ can be split into two terms,
\begin{equation}
\Delta(\tau)\,\psi_m^F(\tau,z) =  \varphi^\mathrm{mock}_{2,m}(\tau,z) +\varphi^\mathrm{true}_{2,m}(\tau,z) \, , 
\end{equation}
whose defining characteristics are as follows. The first term reflects the mock character of the counting function, and as such encodes the failure of $\psi_m^F$ to be modular, while the second is a weakly holomorphic Jacobi form. Adding any (weak) Jacobi form of weight 2 and index $m$ to $\varphi_{2,m}^{\mathrm{mock}}$ and subtracting it from $\varphi_{2,m}^{\mathrm{true}}$ does not modify the defining properties of these functions. Therefore, the explicit form of $\varphi_{2,m}^{\mathrm{mock}}$ is obtained only after imposing another condition. Interestingly, the splitting ambiguity is (almost) removed\footnote{The splitting becomes more subtle and the choice ceases to be unique when the index $m$ is not a prime power. See below for an explicit example.} by demanding an \emph{optimal} growth condition on the Fourier coefficients of $\varphi^{\mathrm{mock}}_{2,m}$, which translates in a growth of at most $\,\exp(\frac{\pi}{m}\sqrt{4mn-\ell^2})$.\\ \\
Based on the above considerations, we define $\Phi_{2,m}^{\mathrm{opt}}$ to be the mock Jacobi form obtained from $\varphi^{\mathrm{mock}}_{2,m}$ after imposing the optimal growth condition. So long as $m$ is a prime power, the optimal mock Jacobi form $\Phi_{2,m}^{\mathrm{opt}}$ is unique \cite{Dabholkar:2012nd}, and it can be expressed in terms of the Hurwitz-Kronecker class numbers\footnote{Hurwitz-Kronecker class numbers are defined as the number of $PSL(2,\mathbb{Z})$-equivalent classes of quadratic forms of discriminant $-n$, weighted by the inverse of the order of their stabilizer in $PSL(2,\mathbb{Z})$.}
\begin{equation}
\label{eq:opt-prime-power}
\Phi_{2,m}^{\mathrm{opt}}(\tau,z) = \mathcal{H}(\tau,z)|\mathcal{V}^{(1)}_{2,m} \, , \quad \mathrm{for} \;\, m \;\, \textnormal{a prime power} \, .
\end{equation}
The optimality in this case translates into the choice of a holomorphic mock Jacobi form. The action of the Hecke-like operator $\mathcal{V}^{(1)}_{k,m}$ is defined in \eqref{eq:app-Hecke}, while $\mathcal{H}(\tau,z)$ is the mock Jacobi form of weight 2 and index 1 whose Fourier coefficients are given by Hurwitz-Kronecker numbers,
\begin{equation}
\mathcal{H}(\tau,z) = \mathcal{H}_0(\tau)\,\vartheta_{1,0}(\tau,z) + \mathcal{H}_1(\tau)\,\vartheta_{1,1}(\tau,z) \, , 
\end{equation}
with
\begin{equation}
\mathcal{H}_\ell(\tau) = \sum_{n=0}^{\infty} H(4n+3\ell)\,q^{n+\frac{3\ell}{4}} \, , \qquad \ell=\{0,1\} \, ,
\end{equation}
and $\vartheta_{m,\ell}(\tau,z)$ is defined in \eqref{eq:app-vartheta}. The first few Hurwitz-Kronecker numbers are given in Table \ref{tab:HK-numbers}. By convention we have the value $H(0) = -1/12$, and $H(n) = 0$ for $n < 0$.

\begin{table}
\centering
\begin{tabular}{c|cccccccccccccccc}
$n$ & 3 & 4 & 7 & 8 & 11 & 12 & 15 & 16 & 19 & 20 & 23 & 24 & 27 & 28 & 31 & 32 \\ 
\hline \vspace{-2.5mm} \\ 
$H(n)$ & 1/3 & 1/2 & 1 & 1 & 1 & 4/3 & 2 & 3/2 & 1 & 2 & 3 & 2 & 4/3 & 2 & 3 & 2
\end{tabular}
\caption{\small The Hurwitz-Kronecker numbers for the first few values of $n$. \label{tab:HK-numbers}}
\end{table}

\noindent When $m$ is not a prime power, the choice of an optimal function is not unique \cite{Dabholkar:2012nd}. For instance, there are three choices of optimal functions for $m=6$, the first non-prime power: 
\begin{align}
\label{eq:opt-6}
\Phi_{2,6}^{\mathrm{opt},\,\mathrm{I}}(\tau,z) =&\, \frac12\mathcal{H}|\mathcal{V}^{(1)}_{2,6}(\tau,z) + \frac1{24}\,\mathcal{F}_6(\tau,z) \, , \\
\Phi_{2,6}^{\mathrm{opt},\,\mathrm{II}}(\tau,z) =&\, \Phi_{2,6}^{\mathrm{opt},\,\mathrm{I}}(\tau,z) - \frac1{24}\,\mathcal{K}_6(\tau,z) \, , \\
\Phi_{2,6}^{\mathrm{opt},\,\mathrm{III}}(\tau,z) =&\, \Phi_{2,6}^{\mathrm{opt},\,\mathrm{I}}(\tau,z) + \frac14\,\mathcal{K}_6(\tau,z) \, ,
\end{align}
where the mock Jacobi forms of weight 2 and index 6 $\mathcal{F}_6(\tau,z)$ and $\mathcal{K}_6(\tau,z)$ are given in Appendix \ref{sec:app-special-Jacobi}. In the following, we will deal explicitly (see footnote \ref{foot:numerics}) with the cases $m=1 \ldots 7$, to incorporate cases where the index is prime (1, 2, 3, 5 and 7), a prime power (4), or neither (6). However, the extension to other non-prime power cases can also be obtained using results contained in \cite{Dabholkar:2012nd}.\\ 
For any index, the completion of $\Phi_{2,\,m}^{\mathrm{opt}}$, which is a Jacobi form of weight 2 and index $m$, satisfies \cite{Dabholkar:2012nd}
\begin{equation}
\label{eq:DMZ-9.5}
\widehat{\Phi}_{2,\,m}^{\mathrm{opt}}(\tau,z) = \Phi_{2,\,m}^{\mathrm{opt}}(\tau,z) - \sqrt{\frac{m}{4\pi}}\sum_{\ell\in\mathbb{Z}/2m\mathbb{Z}}(\vartheta_{m,\ell}^0)^*(\tau)\vartheta_{m,\ell}(\tau,z) \, , 
\end{equation}
where $\vartheta_{m,\ell}^0(\tau) := \vartheta_{m,\ell}(\tau,z)|_{z=0}$ and ${}^*$ denotes the Eichler integral \eqref{eq:Eichler}. Therefore, the theta-decomposition of $\Phi_{2,\,m}^{\mathrm{opt}}$,
\begin{equation}
\label{eq:opt-theta-decomp}
\Phi_{2,\,m}^{\mathrm{opt}}(\tau,z) = \sum_{\ell \in \mathbb{Z}/2m\mathbb{Z}} h_\ell^{\mathrm{opt}}(\tau)\,\vartheta_{m,\ell}(\tau,z) \, , 
\end{equation}
specifies the optimal mock modular forms\footnote{The components of the vector-valued mock modular form $\vec{h}^{\mathrm{opt}}(\tau)$ satisfy $h_\ell^{\mathrm{opt}}(\tau)\, = h_{-\ell}^{\mathrm{opt}}(\tau)$, due to the symmetries of the theta function together with the modular properties of the Jacobi form $\widehat{\Phi}_{2,\,m}^{\mathrm{opt}}$.} $h_\ell^{\mathrm{opt}}$ of weight 3/2 and shadow given by the unary theta series of weight 1/2, $\vartheta_{m,\ell}^0(\tau)$. This suffices to show that, for any $m$, the completion of $h_\ell^{\mathrm{opt}}$, $\widehat{h}_{\ell}^{\mathrm{opt}}$, is a modular form of weight 3/2 with a multiplier system dual to that of $\vartheta_{m,\ell}^0(\tau)$. This multiplier system is discussed in Appendix \ref{sec:app-Jac}.\\ \\
Eventually, the Fourier coefficients of the mock Jacobi forms $\psi_m^F$ can be determined provided one can compute the coefficients of the vector-valued {mixed} mock modular forms entering their theta-decomposition,
\begin{equation}
\label{eq:psimF-split}
\psi_m^F(\tau,z) = \frac{\varphi^\mathrm{true}_{2,m}(\tau,z)}{\eta^{24}(\tau)} + \frac{\Phi^\mathrm{opt}_{2,m}(\tau,z)}{\eta^{24}(\tau)} = \sum_{\ell\in\mathbb{Z}/2m\mathbb{Z}}\,\Biggl[\frac{h^\mathrm{true}_\ell(\tau)}{\eta^{24}(\tau)} + \frac{h^\mathrm{opt}_\ell(\tau)}{\eta^{24}(\tau)}\Biggr]\,\vartheta_{m,\ell}(\tau,z) \, .
\end{equation} 
In particular, this structure shows that the standard Rademacher expansion can be applied to the first term in the theta-decomposition, which is nothing but a modular form, and that obtaining the coefficients of $\psi_m^F$ reduces to finding an expression for the coefficients of ${h^\mathrm{opt}_\ell(\tau)}/{\eta^{24}(\tau)}$. The latter are vector-valued mixed mock modular forms of weight $-21/2$ and dimension $2m$. Following the treatment of Bringmann and Manschot \cite{Bringmann:2010sd}, we can generalize the circle method to recover the Fourier coefficients of such forms. Focusing on the mixed mock modular part, we define
\begin{equation}
\label{eqn:ourfct}
f_{m,\ell}(\tau) := \frac{h_\ell^\mathrm{opt}(\tau)}{\eta^{24}(\tau)} = \sum_{n \geq n_0}\,\alpha_m(n,\ell)\,q^{n-\frac{\ell^2}{4m}} \, , \quad \ell \in \mathbb{Z}/2m\mathbb{Z} \, ,
\end{equation}
where $n_0 = 0$ for $\ell \neq 0$ and $n_0 = -1$ for $\ell = 0$, and
\begin{equation}
\widetilde{f}_{m,\ell}(\tau) := q^{\frac{\ell^2}{4m}}\,f_{m,\ell}(\tau) \, .
\end{equation}
Applying Cauchy's theorem to $\widetilde{f}_{m,\ell}$ and decomposing the contour integral via the Farey sequence as explained in section \ref{sec:circle-method}, we arrive at an equation similar to \eqref{eqn:circ3}, the major difference being that $\widetilde{f}_{m,\ell}$ is now a component of a vector-valued form, 
\begin{equation}
\label{eqn:circ31}
\alpha_m(n,\ell) = \sum_{\substack{0\leq h<k\leq N \\ (h,k)=1}} e^{-2i\pi n \frac{h}{k}} \int_{-\vartheta'_{h,k}}^{\vartheta''_{h,k}}\,\widetilde{f}_{m,\ell}\Bigl(e^{-\frac{2\pi}{N^2} + 2i\pi\frac{h}{k} + 2i\pi\phi}\Bigr)e^{\frac{2\pi n}{N^2} - 2i\pi n\phi}\,\mathrm{d}\phi \, .
\end{equation}
For the time being, we restrict to the Fourier coefficients with $4mn-\ell^2 > 0$. We introduce the variable $z = (\frac{k}{N^2} - i k \phi)$ and use the modular property of $f_{m,\ell}$. The form of the latter is dictated by the functional equation 
\begin{align}
\label{eq:f-transfo}
f_{m,\ell}\Bigl(\frac1k(h+iz)\Bigr) =&\, z^{21/2}\,\psi(\gamma)_{\ell j}\,f_{m,j}\Bigl(\frac1k\Bigl(h'+\frac{i}{z}\Bigr)\Bigr) \nonumber \\
&\,- \sqrt{\frac{m}{8\pi^2}}\,z^{21/2}\,\psi(\gamma)_{\ell j}\,\eta^{-24}\Bigl(\frac1k\Bigl(h'+\frac{i}{z}\Bigr)\Bigr)\,\mathcal{I}_{m,j}\Bigl(\frac{1}{kz}\Bigr) \, , 
\end{align}
where a sum over $j\in\mathbb{Z}/2m\mathbb{Z}$ is implied, the multiplier system $\psi(\gamma)$ is given in \eqref{eq:app-psi-mult} and
\begin{equation}
\label{eq:integral}
\mathcal{I}_{m,j}(x) = \int_0^\infty \frac{\vartheta^0_{m,j}(iw-\frac{h'}{k})}{(w+x)^{3/2}}\,\mathrm{d}w \, .
\end{equation}
Due to the mixed-mock character of $f_{m,\ell}$, the modular transformation is contaminated by the shadow of $\Phi^\mathrm{opt}_{2,m}$ (the Eichler integral of the unary theta series as in \eqref{eq:DMZ-9.5}) divided by the discriminant function. Therefore, using this transformation rule,
\begin{align}
\label{eq:alpha-N}
\alpha_m(n,\ell) =&\, \sum_{\substack{0 \leq h < k \leq N \\ (h,k) = 1}} z^{21/2}\,e^{-2i\pi(n-\frac{\ell^2}{4m})\frac{h}{k}}  \int_{-\vartheta_{h,k}'}^{\vartheta_{h,k}''} \, e^{\frac{2\pi}{k}(n-\frac{\ell^2}{4m})z}\,\psi(\gamma)_{\ell j}\,\times \nonumber \\
&\,\Biggl[f_{m,j}\Bigl(\frac1k\Bigl(h'+\frac{i}{z}\Bigr)\Bigr) - \sqrt{\frac{m}{8\pi^2}}\,\eta^{-24}\Bigl(\frac1k\Bigl(h'+\frac{i}{z}\Bigr)\Bigr)\,\mathcal{I}_{m,j}\Bigl(\frac{1}{kz}\Bigr) \Biggr] \mathrm{d}\phi\, ,
\end{align}
where again a sum over $j\in \mathbb{Z}/2m\mathbb{Z}$ is implicit. 
Two distinct terms appear in the integrand: one reflects modularity while the other is generated solely by the shadow of $h_\ell^{\mathrm{opt}}$ through the $\mathcal{I}_{m,j}$ integral. We denote these terms by $\Sigma_1$ and $\Sigma_2$, respectively. Note that up to now, we haven't made use of the explicit form of the function $f_{m,\ell}$ but only of its transformation properties.\\
To complete the circle method, one has to estimate the behavior of the different terms in the limit for $N \rightarrow \infty$ and thus prove the convergence of the above series. The analysis yielding the exact expression of the Fourier coefficients $\alpha_m(n,\ell)$ is performed in Appendix \ref{sec:app-asymptotic}, and here we only mention the main steps of the proof.\\ 
The first term $\Sigma_1$ in \eqref{eq:alpha-N} is dominated (in the limit $N \rightarrow \infty$) by the polar terms of $f_{m,\ell}$. We denote their contribution by $\Sigma_1^*$ and refer to them using tilde variables. Introducing the usual combinations
\begin{equation}
\Delta := 4mn-\ell^2 \, , \qquad \widetilde{\Delta} := 4m\widetilde{n}-\widetilde{\ell}^{\,2} \, ,
\end{equation}
as well as $\widetilde{n}_0$ such that 
\begin{equation}
\label{eq:n0}
\widetilde{n}_0 = \begin{cases} 0 \quad\; \mathrm{for} \;\; \widetilde{\ell} \neq 0 \\ -1 \;\; \mathrm{for} \;\; \widetilde{\ell} = 0 \, , \end{cases}
\end{equation}
we have
\begin{equation}
\label{eq:sigma1star}
\Sigma_1^* = \sum_{\substack{\widetilde{n} \geq \widetilde{n}_0 \\ \widetilde{\ell} \in \mathbb{Z}/2m\mathbb{Z} \\ \widetilde{\Delta} < 0}} \alpha_m(\widetilde{n},\widetilde{\ell})\sum_{\substack{0 \leq h < k \leq N \\ (h,k) = 1}}e^{2\pi i \bigl(-\tfrac{h}{k}\frac{\Delta}{4m} + \tfrac{h'}{k}\frac{\widetilde{\Delta}}{4m}\bigr)} \psi(\gamma)_{\ell\widetilde{\ell}}\int_{-\vartheta_{h,k}'}^{\vartheta_{h,k}''} z^{21/2}e^{\tfrac{2\pi}{k}\bigl(z\frac{\Delta}{4m}-\frac{\widetilde{\Delta}}{4mz}\bigr)}\;\mathrm{d}\phi \, .
\end{equation}
Here, $\alpha_m(\widetilde{n},\widetilde{\ell})$ are the polar coefficients of $f_{m,\ell}$. Using \eqref{eq:opt-prime-power} and taking into account the discriminant function in the denominator, we have the following analytic expression in terms of the Hurwitz-Kronecker numbers\footnote{Recall that by convention $H(n) = 0$ for $n<0$.}  displayed in Table \ref{tab:HK-numbers},
\begin{align}
\alpha_m(\widetilde{n},\widetilde{\ell}) =&\, \sum_{d|(\widetilde{n}+1,\widetilde{\ell},m)} d\,H\Bigl(\frac{4m(\widetilde{n}+1) - \widetilde{\ell}^{\,2}}{d^2}\Bigr) \qquad \qquad \mathrm{for} \;\, m \;\, \textnormal{prime}, \\
\alpha_4(\widetilde{n},\widetilde{\ell}) =&\, \sum_{d|(\widetilde{n}+1,\widetilde{\ell},4)} d\,H\Bigl(\frac{16(\widetilde{n}+1) - \widetilde{\ell}^{\,2}}{d^2}\Bigr) - 2\begin{cases}H\bigl(4(\widetilde{n}+1) - \bigl(\frac{\widetilde{\ell}}{2}\,\bigr)^2\,\bigr) \;\; \mathrm{if} \;\; 2 | \widetilde{\ell} \, , \\ \qquad \qquad 0 \qquad \qquad \mathrm{otherwise}\, .\end{cases}
\end{align}
The case where $m=6$ needs to be treated separately since, as we mentioned above, the choice of an optimal function is not unique. If we make the choice (I) in \eqref{eq:opt-6}, then
\begin{equation}
\alpha_6(\widetilde{n},\widetilde{\ell}) = \frac12\,\sum_{d|(\widetilde{n}+1,\widetilde{\ell},6)} d\,H\Bigl(\frac{24(\widetilde{n}+1) - \widetilde{\ell}^{\,2}}{d^2}\Bigr) + \frac1{24}\,c_{\mathcal{F}_6}(\widetilde{n},\widetilde{\ell}) \, , 
\end{equation}
where $c_{\mathcal{F}_6}$ are the polar coefficients of the mock Jacobi form $\mathcal{F}_6$ \eqref{eq:app-F6}, given by
\begin{align}
c_{\mathcal{F}_6}(-1,1) =&\, c_{\mathcal{F}_6}(4,11) = -c_{\mathcal{F}_6}(0,5) = -c_{\mathcal{F}_6}(1,7) = -1 \, , \\
c_{\mathcal{F}_6}(0,1) =&\, c_{\mathcal{F}_6}(5,11) = -c_{\mathcal{F}_6}(1,5) = -c_{\mathcal{F}_6}(2,7) = 11 \, .
\end{align}
Similar formulae for the polar coefficients in the case where $m$ is a prime power greater than 4 or $m$ not a prime power can also be obtained by applying the definitions of the optimal mock modular forms \eqref{eq:opt-prime-power} and the results of \cite{Dabholkar:2012nd}. \\ \\
Following \cite{Bringmann:2010sd}, the boundaries of the integral in \eqref{eq:sigma1star} can be written in a symmetric form up to an error term (see Appendix \ref{sec:app-asymptotic}) which vanishes in the $N \rightarrow \infty$ limit. We are then left with the integral representation of the standard I-Bessel function of weight 23/2 \eqref{eq:app-Bessel}. If we now define
\begin{equation}
\label{eq:Klooster}
Kl(\mu,\nu\,;k,\psi)_{ij} := \sum_{\substack{0\leq h < k \\ (h,k) = 1}}\, e^{2\pi i \bigl(-\frac{h}{k}\mu + \frac{h'}{k}{\nu}\bigr)}\,\psi(\gamma)_{ij} \, ,
\end{equation}
we obtain a first Rademacher-type contribution to the coefficients $\alpha_m(n,\ell)$,
\begin{equation}
\Sigma_1 = \sum_{\substack{\widetilde{n} \geq  \widetilde{n}_0 \\ \widetilde{\ell} \in \mathbb{Z}/2m\mathbb{Z} \\ \widetilde{\Delta} < 0}} \alpha_m(\widetilde{n}, \widetilde{\ell})\, \sum_{k=1}^\infty\,Kl\Bigl(\frac{\Delta}{4m},\frac{\widetilde{\Delta}}{4m}\,;k,\psi\Bigr)_{\ell\widetilde{\ell}}\,\frac{2\pi}{k}\biggl(\frac{|\widetilde{\Delta}|}{\Delta}\biggr)^{23/4}\,I_{23/2}\biggl(\frac{\pi}{m k}\sqrt{|\widetilde{\Delta}|\Delta}\biggr)  \, .
\end{equation}
\noindent We now have to deal with the shadow contribution $\Sigma_2$ in \eqref{eq:alpha-N}, which needs a bit more care. Implementing the results of Appendix \ref{sec:app-asymptotic} regarding the representation of the Eichler integral of $\vartheta^{0}_{m,j}(\tau)$, we obtain a more suitable form for the estimation of the asymptotic of the latter via Mittag-Leffler theory:
\be
\int_0^{\infty}\frac{\vartheta^0_{m,j} (iw-\frac{h'}{k})}{(w+x)^{3/2}} \mathrm{d}w= \sum\limits_{\substack{g\,(2mk) \\ g\equiv j(2m)}}\,e^{-\pi i \frac{g^2 h'}{2mk}}\biggl(\frac{2 \delta_{0,g}}{\sqrt{x}} - \frac{1}{\sqrt{2m}\,\pi k^2 x}\int_{-\infty}^{+\infty}e^{-2\pi x{m}u^2}f_{k,g,m}(u)\,\mathrm{d}u\biggr) \, ,
\ee
\be
\label{eq:fkgmu}
f_{k,g,m}(u):=\begin{cases} \frac{\pi^2}{\text{sinh}^2(\frac{\pi u}{k}-\frac{\pi i g}{2mk})} & \text{if} \, g\not\equiv 0 \,( 2mk) \, , \\ \frac{\pi^2}{\text{sinh}^2(\frac{\pi u}{k})}-\frac{k^2}{u^2} & \text{if} \, g\equiv 0 \,( 2mk) \, ,
\end{cases} 
\ee
where $\delta_{0,g} = 0$ unless $g\equiv 0\;(2mk)$ in which case it is equal to one\footnote{To shorten the notation from now on, we will adopt the convention of writing only brackets when the variable is defined over a finite field. For instance, $g\equiv j \,(\text{mod} \, 2m)$ becomes $g\equiv j \,( 2m)$ and $g\,(2mk)$ stands for $g\in \mathbb{Z}/2mk\mathbb{Z}$.}. This shows that there are two contributions to $\Sigma_2$, one coming from $g\equiv 0\,(2mk)$ and the other from $g \not\equiv 0\,(2mk)$. For both of them, only the polar coefficient of $\eta(\tau)^{-24}$ contributes to the $N \rightarrow \infty$ limit. After evaluating the integrals over the Farey sequences in the same way as for $\Sigma_1$, we are thus left with two contributions
\begin{equation}
\Sigma_{2,\,g\equiv0(2mk)} = \frac{\sqrt{2m}}{\kappa}\,\sum_{k=1}^\infty Kl\Bigl(\frac{\Delta}{4m},-1\,;k,\psi\Bigr)_{\ell 0}\frac{1}{\sqrt{k}}\biggl(\frac{4m}{\Delta}\biggr)^{6}\,I_{12}\biggl(\frac{2\pi}{k\sqrt{m}}\sqrt{\Delta}\biggr) \, ,
\end{equation}
and
\begin{align}
\Sigma_{2,\,g \not\equiv 0(2mk)} =&\, -\frac{1}{2\pi \kappa}\,\sum_{k=1}^\infty\,\sum_{\substack{j\in\mathbb{Z}/2m\mathbb{Z} \\ g(2mk) \\ g \equiv j (2m)}}\,Kl\Bigl(\frac{\Delta}{4m},-1-\frac{g^2}{4m}\,;k,\psi\Bigr)_{\ell j}\frac{1}{k^2}\biggl(\frac{4m}{\Delta}\biggr)^{25/4}\,\times \nonumber \\
&\int_{-1/\sqrt{m}}^{+1/\sqrt{m}}\,f_{k,g,m}(u)\,I_{25/2}\Biggl(\frac{2\pi}{k\sqrt{m}}\sqrt{\Delta(1-m u^2)}\Biggr)(1-m u^2)^{25/4}\,\mathrm{d}u \, .
\end{align}
The normalization factor $\kappa$ in the two expressions above is 2 for $m=1$ and 1 otherwise, and is related to the normalization of the Eichler integral of the shadow in \eqref{eq:DMZ-9.5}. It has been chosen to be consistent with the $m=1$ case discussed in \cite{Bringmann:2010sd}.\\ \\
Putting all three contributions together, we arrive at the following formula\footnote{Numerically this formula has been tested up to $m=7$, in order to have a direct comparison with the supergravity computation presented in \cite{Murthy:2015zzy} and to include the cases where $m$ is prime, a prime power or neither. However the proof applies for any index.\label{foot:numerics}} for the Fourier coefficients of $f_{m,\ell}$ when $\Delta = 4mn-\ell^2 >0$,

\begin{align}
\label{eq:mixed-mock-coeffs}
\alpha_m(n,\ell) =&\, 2\pi\sum_{\substack{\widetilde{n} \geq \widetilde{n}_0 \\ \widetilde{\ell} \in \mathbb{Z}/2m\mathbb{Z} \\ \widetilde{\Delta} < 0}} \alpha_m(\widetilde{n},\widetilde{\ell})\, \sum_{k=1}^\infty\frac{Kl\bigl(\frac{\Delta}{4m},\frac{\widetilde{\Delta}}{4m}\,;k,\psi\bigr)_{\ell\widetilde{\ell}}}{k}\biggl(\frac{|\widetilde{\Delta}|}{\Delta}\biggr)^{23/4}\,I_{23/2}\biggl(\frac{\pi}{m k}\sqrt{|\widetilde{\Delta}|\Delta}\biggr) \nonumber \\[1mm]
&\,+ \frac{\sqrt{2m}}{\kappa}\sum_{k=1}^\infty \frac{Kl\bigl(\frac{\Delta}{4m},-1\,;k,\psi\bigr)_{\ell 0}}{\sqrt{k}}\biggl(\frac{4m}{\Delta}\biggr)^6\,I_{12}\biggl(\frac{2\pi}{k\sqrt{m}}\sqrt{\Delta}\biggr) \\[1mm]
&\,-\frac{1}{2\pi\kappa}\,\sum_{k=1}^\infty\,\sum_{\substack{j\in\mathbb{Z}/2m\mathbb{Z} \\ g(2mk) \\ g \equiv j (2m)}} \frac{Kl\bigl(\frac{\Delta}{4m},-1-\frac{g^2}{4m}\,;k,\psi\bigr)_{\ell j}}{k^2}\biggl(\frac{4m}{\Delta}\biggr)^{25/4}\,\times \nonumber \\
&\qquad\quad\;\; \times \int_{-1/\sqrt{m}}^{+1/\sqrt{m}}\,f_{k,g,m}(u)\,I_{25/2}\Biggl(\frac{2\pi}{k\sqrt{m}}\sqrt{\Delta(1-m u^2)}\Biggr)(1-m u^2)^{25/4}\,\mathrm{d}u \, . \nonumber
\end{align}
\noindent Despite the fact that we derived the above expression for $\Delta> 0$, the limit for $\Delta\rightarrow 0$ exists and it correctly reproduces the constant terms of the vector-valued mock modular forms.\\ \\
It is interesting to examine the $n \rightarrow \infty$ asymptotic behavior of the coefficients $\alpha_m(n,\ell)$. Using the asymptotic behavior of the I-Bessel function in this regime \eqref{eq:app-Bessel-asymptotic}, we find, for $m$ prime or a prime power,
\begin{equation}
\alpha_m(n,\ell) \underset{n\rightarrow\infty}{\sim} \Biggl(-\frac{1}{2\sqrt{m}}\,\alpha_m(-1,0)\,n^{-6} - \frac{1}{2\sqrt{2}\,\kappa\,\pi}\,n^{-25/4} + \mathcal{O}(n^{-13/2})\Biggr)e^{4\pi\sqrt{n}} \, ,
\end{equation}
while for $m=6$ we have
\begin{equation}
\alpha_6(n,\ell) \underset{n\rightarrow\infty}{\sim} \Biggl(-\frac{1}{2\sqrt{6}}\,\beta_\ell\,n^{-6} - \frac{1}{2\sqrt{2}\,\pi}\,n^{-25/4} + \mathcal{O}(n^{-13/2})\Biggr)e^{4\pi\sqrt{n}} \, ,
\end{equation}
with $\beta_\ell = Kl(\infty,-25/24;\,1,\psi)_{\ell 1}$, which is finite and smaller than 1 for $\ell \in \mathbb{Z}/12\mathbb{Z}$. Note that the exponential behavior of the Fourier coefficients is compatible with the result of \cite{Dabholkar:2012nd}, Theorem 9.3, after taking into account the fact that we have divided the optimal mock Jacobi form of weight 2 and index $m$, $\Phi^{\mathrm{opt}}_{2,m}$ in \eqref{eq:opt-prime-power} and \eqref{eq:opt-6}, by the discriminant function.

\section{Comparison with supergravity}
\label{sec:sugra-comp}

As explained below \eqref{eq:psimF-split}, the mock modular forms entering the theta-decomposition of $\psi_m^F$ consist of a truly modular part and a mixed mock modular part. 
Equation \eqref{eq:mixed-mock-coeffs} provides an exact expression for the Fourier coefficients of the latter. To reconstruct the coefficients of the former it is enough to apply the Rademacher expansion \eqref{eq:app-Rad} to the modular form $h_\ell^\mathrm{true}/\eta^{24}$ of weight $-21/2$. \\
As a result, the complete formula for the Fourier coefficients of $\psi_m^F$ contains three types of terms: an $m$-dependent number of I-Bessel functions of weight 23/2 {(coming from both the true and the mock parts)}, an I-Bessel function of weight 12 and the integral of an I-Bessel function of weight 25/2 times a hyperbolic function. The question we would like to address now, harking back to Section \ref{sec:QEF}, is whether an $\mathcal{N}=4$ supergravity computation can reproduce this structure for the degeneracies of single-center (immortal) 1/4-BPS black holes.\\ \\
It was shown in \cite{Murthy:2015zzy} that localizing the quantum entropy function with the prepotential \eqref{eq:prepot-K3} yields a sum of I-Bessel functions of weight 23/2, along with additional contributions that were dubbed ``edge-effects''. A comparison in the I-Bessel 23/2 sector was conducted for the cases $m=1\ldots 7$, and although there seemed to be a good agreement between the two calculations, some discrepancies remained (see Tables in \cite{Murthy:2015zzy}). \\
Rather encouragingly, edge-effects in the supergravity computation were shown to yield I-Bessel functions of weight 12. However, what was obtained is an infinite sum\footnote{The infinite sum arose because the instanton correction proportional to $\widehat{A}$ in the prepotential \eqref{eq:prepot-K3} was written using a Fourier expansion and integrated term by term. } of such Bessel functions\footnote{The precise contribution is recalled in Appendix \ref{sec:app-Iu}, where some errors in \cite{Murthy:2015zzy} have been corrected.}, while the Fourier coefficients of $\psi_m^F$ contain only a single I-Bessel of weight 12 entering via \eqref{eq:mixed-mock-coeffs}. \\
In addition, the measure of the localizing manifold $[\mathrm{d}\phi^I]$ used in \cite{Murthy:2015zzy} was inspired by a saddle-point approximation of the Igusa cusp form derived in \cite{David:2006yn}. 
Using such a set-up, the localized quantum entropy function takes the form \cite{Murthy:2015zzy}, 
\begin{align}
\label{eq:QEF-perfect-square}
\widehat{W}(n,\ell,m) =&\, N_0\sum_{p,\bar{p}\,\geq\,-1}(m - p - \bar{p})\,d(p)\,d(\bar{p})\,e^{i\pi(p-\bar{p})\frac{\ell}{m}} \; \times \\
&\!\!\!\!\!\!\!\!\!\!\!\!\!\!\!\!\!\!\!\!\!\!\!\!\!\!\!\!\!\!\!\!\!\!\!\!\int_{\gamma_2}\frac{\mathrm{d}\tau_2}{\tau_2^{13}} \exp\Bigl[-\pi\tau_2 \frac{\Delta(p,\bar{p})}{m} + \frac{\pi}{\tau_2}\bigl(n - \frac{\ell^2}{4m}\bigr)\Bigr]\int_{\gamma_1} \mathrm{d}\tau_1 \exp\Biggl[\frac{\pi m}{\tau_2}\Bigl(\tau_1 + i(p-\bar{p})\frac{\tau_2}{m}-\frac{\ell}{2m}\Bigr)^2\Biggr] \, , \nonumber
\end{align}
with $\Delta(p,\bar{p}) = 4\,m\,\bar{p} - (m-p + \bar{p})^2$, $d(p)$ the $p^\mathrm{th}$ Fourier coefficient of $\eta(\tau)^{-24}$ ($p$ denotes the instanton number) and $N_0$ a normalization equal to $2^{-12}$. It is important to stress that this localization computation does not include sub-leading saddle-points of the quantum entropy function \eqref{eq:QEF} corresponding to orbifolded near-horizon geometries \cite{Murthy:2009dq}, and it should therefore only be compared to the $k=1$ term in the asymptotic expansion for the Fourier coefficients of the single-center counting function $\psi_m^F$.\\ \\
The next step to compute the integrals deals with the choice of contours $\gamma_1,\,\gamma_2$ (see also \cite{Gomes:2015xcf} for a discussion of such contours). The choice made in \cite{Murthy:2015zzy} led to the correct number of I-Bessel functions of weight 23/2 for any $m$ by limiting the sum in \eqref{eq:QEF-perfect-square} over a finite, $m$-dependent range for $p$ and $\bar{p}$, but introduced the edge-effects. A close look at their contribution reveals that they take the form of I-Bessel functions of weight 12 whose prefactors and arguments are functions of $p$ and $\bar{p}$ but for which the sum is not truncated, as shown in \eqref{eq:app-QEF-Bessel12}. In addition, the integral of the Bessel functions of weight 25/2 multiplied by a hyperbolic function (the $f_{k,g,m}(u)$ defined in \eqref{eq:fkgmu}) in the Fourier coefficients of $\psi_m^F$ seems absent from the supergravity calculation. \\ 
At present, a contour prescription which, while keeping the finite sum over I-Bessel functions of weight 23/2, would also lead to a single I-Bessel function of weight 12 along with an integral of an I-Bessel of weight 25/2 is still missing. It would be illuminating to find motivations based on low-energy physics for this contour and explore whether such considerations could lead to an agreement between the supergravity answer and the Fourier coefficients of $\psi_m^F$ obtained in this paper.\\ \\
To illustrate these discrepancies more concretely, we display the Fourier coefficients of $\psi_1^F$ at $\ell=0$. The first term of the mixed Rademacher series contributes
\begin{align}
k=1:&\;\;-\sqrt{2}\pi\Bigl[3\,\Bigl(\frac{5}{4n}\Bigr)^{23/4}I_{23/2}\bigl(\pi\sqrt{20n}\bigr)- 6\,\Bigl(\frac{1}{n}\Bigr)^{23/4}I_{23/2}\bigl(\pi\sqrt{16n}\bigr) \nonumber \\
&\qquad \quad \;\;\; + 816\,\Bigl(\frac{1}{4n}\Bigr)^{23/4}I_{23/2}\bigl(\pi\sqrt{4n}\bigr)\Bigr] \nonumber \\
&\,-\sqrt{2}\pi\Bigl[54\,\Bigl(\frac{1}{n}\Bigr)^{23/4}I_{23/2}\bigl(\pi\sqrt{16n}\bigr) - 216\,\Bigl(\frac{1}{4n}\Bigr)^{23/4}I_{23/2}\bigl(\pi\sqrt{4n}\bigr)\Bigr] \\
&\,-\frac12\,\Bigl(\frac{1}{n}\Bigr)^6\,I_{12}\bigl(\pi\sqrt{16n}\bigr) \nonumber \\
&\,+\frac{1}{4\sqrt{2}}\Bigl(\frac{1}{n}\Bigr)^{25/4}\int_{-1}^{1}[f_{k,0,1}(u) + f_{k,1,1}(u)]\,I_{25/2}\bigl(\pi\sqrt{16n(1-u^2)}\bigr)(1-u^2)^{25/4}\,\mathrm{d}u \, , \nonumber
\end{align}
where the first bracket are the terms arising from the true modular piece $h_\ell^\mathrm{true}/\eta^{24}$ and the rest are the mixed mock contributions \eqref{eq:mixed-mock-coeffs}. The leading Bessel is unaffected by the latter, but the two pieces start mixing at sub-leading order in $n$. The Bessel functions of weight 23/2 appearing at index one are correctly reproduced, including the prefactors\footnote{For index $m\geq3$, the prefactors are not all reproduced accurately as evidenced by the bold-faced and boxed entries in the Tables of \cite{Murthy:2015zzy}.}, by the supergravity calculation \cite{Murthy:2015zzy}. However, the Bessel function of weight 12 makes a contribution of the same order as that of the first sub-leading Bessel function of weight 23/2 to the entropy of 1/4-BPS black holes. This contribution does not match the one coming from the edge-effects \eqref{eq:app-QEF-Bessel12},
\begin{equation}
\textnormal{Edge-effects}:\qquad -2\sum_{\substack{p\geq-1 \\ p\neq 0}}\,\frac{p-2}{p}\,d(p)\,\Bigl(\frac{1}{n}\Bigr)^6\,I_{12}\bigl(\pi\sqrt{16n}\bigr) \, ,
\end{equation}
since the summand of the prefactor is a monotonically increasing function in $p$, leading to a divergence. Note that the above observations are independent of the higher-$k$ terms in the Rademacher series, since they contribute at most a Bessel function of weight 23/2 with an argument $\pi\sqrt{20n/4}$, which is sub-leading with respect to the Bessel function of weight 12 above. This phenomenon occurs also for higher $m$.\\ 
In the conclusion, we will outline what we believe are necessary modifications to the supergravity calculation which could lead to an agreement between the microscopic and the macroscopic result. 

\section{Conclusion}
\label{sec:conclusion}
In this paper, we proved an exact formula for the Fourier coefficients of the mixed mock modular forms entering the theta-decomposition of the counting function of single-centered 1/4 BPS black holes in $\mathcal{N}=4$ supergravity. This result was obtained using an extension of the circle method first implemented in the context of moduli spaces of stable coherent sheaves on $\mathbb{P}^2$ \cite{Bringmann:2010sd}. The (mock) modular properties of the functions highly constrain the form of the Fourier coefficients, allowing to predict not only the growth of the black hole degeneracies for large charges, but also the precise value of the degeneracy at fixed charges, which receives both perturbative and non-perturbative corrections.\\
Motivated by the precise match between the microstate counting function of 1/8-BPS states in Type IIB compactified on $T^6$ and the quantum entropy function computed by means of localization in the corresponding supergravity theory, we are led to a similar discussion for the case of immortal dyons in the $\mathcal{N}=4$ theories. 
However, as explained in section \ref{sec:sugra-comp}, the low-energy effective field theory computation needs further corrections in order to reproduce the first ($k=1$) term in the Rademacher expansion. Here we mention various subtleties arising in the calculation of the localized quantum entropy function in supergravity.\\
The supergravity localization result relies on the form of the measure $[\mathrm{d}\phi^I]$ along the localizing manifold. Such a measure is difficult to obtain from first principles. By definition, it is the induced measure from the full field configuration space of $\mathcal{N}=2$ superconformal gravity to the localizing manifold (which is a particular slice in the configuration space specified by BPS solutions). However, the former measure is not known at present. 
The form used in \cite{Murthy:2015zzy} was borrowed from a saddle-point approximation of the microscopic degeneracies; in order to obtain the exact measure, one might need to include corrections to this approximation which would lead to a modification of the final supergravity result.\\ 
The choice of contour for the complex integrals \eqref{eq:QEF-perfect-square} could also be modified to exhibit the same structure as the Fourier coefficients of the single-center counting functions $\psi_m^F$. 
Lastly, by inverting the approach, we believe that the exact microscopic results obtained herein could be used to infer what the localizing measure is and what the contour of integration in the localized quantum entropy function must be in order to guarantee a matching between string-theoretic and supergravity counting of 1/4-BPS states in $\mathcal{N}=4$ theories. 
Moreover, the exact expression of the coefficients might shed some light on the type of geometries needed to reproduce the correct sum over saddle-points (near-horizon geometries) in the full $W(q,p)$. \\
In conclusion, our result suggests that some aspects of the supergravity result would benefit from a systematic analysis. Such an analysis should be conducted under the guidance of the exact microscopic results obtained in section \ref{sec:mixed-Rad}, which provides a precise goal to aim for in the low-energy theory. We hope to report on these issues in the future.\\ 
The results of section \ref{sec:mixed-Rad} may also find wider applications to other types of mixed mock modular forms arising in different physical contexts. For instance, our results might be extended to the mixed mock modular forms arising in compactifications of string theory on $\mathrm{CY}_3$--folds leading to four-dimensional black holes with $\mathcal{N}=2$ supersymmetry\cite{Klemm}, or the related five-dimensional spinning black holes \cite{Ftheo}.
\vspace{-3mm}

\section*{Acknowledgements}
The authors are grateful to Miranda Cheng and Sameer Murthy for their invaluable help in putting together the manuscript, as well as for insightful comments. 
We would also like to thank Jan Manschot for clarifications regarding the extension of the circle method and Sam van Leuven, Guglielmo Lockhart, Noppadol Mekareeya and Daniel Whalen for useful discussions. 
The work of VR is supported in part by INFN and by the ERC Starting Grant 637844-HBQFTNCER.


\begin{appendix}

\section{Modular miscellaneous}
\label{app:mod-misc}

\subsection{(Mock) Jacobi forms}
\label{sec:app-Jac}

A Jacobi form\cite{EichZagier} of weight~$w$ and index~$m$ with respect to the fundamental modular group $SL(2,\mathbb{Z})$ is a holomorphic function~$\varphi(\tau,z):\mathbb{H}\times \mathbb{C} \rightarrow \mathbb{C}$ (where~$\mathbb{H}$ is the upper half-plane) which satisfies two functional equations 
\begin{align}
\label{eq:app-mod-transform}  
\varphi\Bigl(\frac{a\tau+b}{c\tau+d},\frac{z}{c\tau+d}\Bigr) =&\, (c\tau+d)^w\,e^{\frac{2\pi imc z^2}{c\tau+d}}\,\varphi(\tau,z)  \qquad \forall \quad \left(\begin{array}{cc} a&b\\ c&d \end{array}\right) \in SL(2,\mathbb{Z}) \, , \\[2mm]
\label{eq:app-elliptic-transfo}
\varphi(\tau, z+\lambda\tau+\mu) =&\, e^{-2\pi i m(\lambda^2 \tau + 2 \lambda z)} \varphi(\tau, z) \qquad \quad \; \forall \quad \lambda,\,\mu \in \mathbb{Z} \, . 
\vspace{.3cm}
\end{align}
Due to the periodicity properties encoded in the above equations, $\varphi(\tau,z)$ has a Fourier expansion 
\begin{equation}
\varphi(\tau,z) = \sum_{n,\ell\in\mathbb{Z}}\,c(n,\ell)\,q^n\,y^\ell \, ,
\end{equation}
where~$q:=e^{2\pi i \tau}$ and~$y:=e^{2\pi i z}$.
Depending on the asymptotic growth of the coefficients, a Jacobi form is called \emph{weak} (when $c(n,\ell) = 0$ unless $n \geq 0$), \emph{holomorphic} (when $c(n,\ell) = 0$ unless $4mn \geq \ell^2$) or a \emph{cusp} form (when $c(n,\ell) = 0$ unless $4mn > \ell^2 $). Lastly, if the coefficients satisfy the weaker condition that $c(n,\ell) = 0$ unless $n \geq n_0$ for some possibly negative integer $n_0$, the associated Jacobi form is called \emph{weakly holomorphic}.\\
The elliptic transformation \eqref{eq:app-elliptic-transfo} implies the following periodicity property of the Fourier coefficients
\begin{equation}
c(n,\ell) = C_{\ell\,}(\Delta) \, , \quad \mathrm{where} \; C_{\ell\,}(\Delta) \; \textnormal{depends only on} \; \ell \; \mathrm{mod} \; 2m \,\, \text{and}\, \,\Delta:= 4mn-\ell^2.
\end{equation}
Also owing to \eqref{eq:app-elliptic-transfo}, a Jacobi form of weight $w$ and index $m$ can be decomposed into a vector-valued modular form of weight $w-1/2$ via its theta-decomposition
\begin{equation}
\label{eq:app-theta-decomp}
\varphi(\tau,z) = \sum_{\ell \in \mathbb{Z}/2m\mathbb{Z}} h_\ell(\tau)\,\vartheta_{m,\ell}(\tau,z) \, ,
\end{equation} 
where the components $h_\ell(\tau)$ take the form 
\begin{equation}
h_\ell(\tau) = \sum_\Delta\,C_{\ell\,}(\Delta)\,q^{\Delta/4m} \, .
\end{equation}
The $\vartheta_{m,\ell}(\tau,z)$ denote the standard weight 1/2, index $m$ theta function, 
\begin{equation}
\label{eq:app-vartheta}
\vartheta_{m,\ell}(\tau,z) := \sum\limits_{\substack{r\in\mathbb{Z}\\ r\equiv \ell \, \text{mod}\, 2m}}\,q^{r^2/4m}\,y^{r} \, ,
\end{equation}
and obey the functional equations
\be
\begin{cases}
\label{eq:thetatransf}
\vartheta_{m,\ell}(-1/\tau,-z/\tau)= e^{2\pi i \frac{mz^2}{\tau}}\,\sqrt{-i\tau}\,\rho(S)_{\ell j}\,\vartheta_{m,j}(\tau,z) \, , \\[2mm]
\vartheta_{m,\ell}(\tau+1,z)= \rho(T)_{\ell j} \,  \vartheta_{m,j}(\tau,z) \, ,
\end{cases}
\ee
where a sum over $j\in\mathbb{Z}/2m\mathbb{Z}$ is implied, and $\rho(S)$ and $\rho(T)$ are $2m$-dimensional matrices defining the multiplier system of $ \vartheta_{m,\ell}(\tau,z)$ \cite{EichZagier} 
\be
\rho(S)_{\ell j}= \frac{1}{\sqrt{2m}}\,e^{-2\pi i \frac{\ell j}{2m}} \, , \qquad \rho(T)_{\ell j}=e^{2\pi i \frac{\ell^2}{4m}}\,\delta_{\ell, j}\,  
\ee
through its action on the $SL(2,\mathbb{Z})$ generators
\begin{equation}
\label{eq:sl2zgen}
S = \begin{pmatrix} 0 & -1 \\ 1 & 0 \end{pmatrix} \, , \qquad T = \begin{pmatrix} 1 & 1 \\ 0 & 1 \end{pmatrix} \, .
\end{equation}
In section \ref{sec:mixed-Rad}, we will also need a related multiplier system,
\begin{equation}
\label{eq:app-psi-mult}
\psi(\gamma)_{\ell j} := -e^{i\pi/4}\,\rho(\gamma)^{-1}_{\ell j} \, ,
\end{equation}
which is defined to act on any $SL(2,\mathbb{Z})$ matrix $\gamma$. The explicit expressions can be obtained by writing $\gamma$ in terms of the generators $S$ and $T$.\\ \\
An extension of the definition of Jacobi forms is provided by \emph{mock Jacobi forms}, whose theta-decomposition yields mock modular forms of weight $w$ and whose completion is
\be
\widehat{\varphi}(\tau,z)= {\varphi}(\tau,z)+ \sum_{\ell \in \mathbb{Z}/2m\mathbb{Z}} g^*_\ell(\tau)\,\vartheta_{m,\ell}(\tau,z) \, ,
\ee
where now $\widehat{\varphi}$ transforms according to \eqref{eq:app-mod-transform}, \eqref{eq:app-elliptic-transfo} with weight $w+1/2$ and index $m$, and the Eichler integral $g_\ell^*(\tau)$ is defined by\footnote{Note that this representation of the Eichler integral is valid for $w>1$ or for $g_\ell(\tau)$ a cusp form.}
\be
g_\ell^*(\tau)= \Bigl(\frac{i}{2\pi}\Bigr)^{w-1}\int_{-\bar{\tau}}^{\,i\infty}\,(z+\tau)^{-w}\,\overline{g_\ell(-\bar{z})}\,\mathrm{d}z.
\ee
Consider now the Rademacher series introduced in section \ref{sec:circle-method}. As mentioned in the main text, so long as $g(\tau)$ is a cusp form, the Rademacher series provides a powerful tool to reconstruct the Fourier coefficients of the mock modular form (and thus of the mock Jacobi form). 
We illustrate here the Rademacher expansion which applies to mock modular forms of weight smaller or equal to zero, modular group $SL(2,\mathbb{Z})$ and generic multiplier system $\psi(\gamma)$ \cite{Whalen}. Once the modular properties of the mock modular forms $h_\ell(\tau)$ are known, the only extra ingredient required to determine the Fourier coefficients $C_{{\ell}\,}({\Delta})$ are the \emph{polar terms}, i.e. the terms with negative powers of $q$ in the Fourier expansion
\begin{equation}
h_\ell(\tau) = \sum_{\widetilde{\Delta} < 0} C_{\widetilde{\ell}\,}(\widetilde{\Delta})\,q^{\widetilde{\Delta}/4m} +\sum_{{\Delta} \ge 0} C_{{\ell}\,}({\Delta})\,q^{{\Delta}/4m}  \, .
\end{equation}
In turn, the Rademacher series for the Fourier coefficients of $h_\ell(\tau)$ takes the form
\begin{equation}
\label{eq:app-Rad}
C_\ell(\Delta) = 2\pi\,\sum_{k=1}^{\infty}\,\sum_{\substack{\widetilde{\ell}\in\mathbb{Z}/2m\mathbb{Z} \\ \widetilde{\Delta} < 0}}\,C_{\widetilde{\ell}\,}(\widetilde{\Delta})\,\frac{Kl\bigl(\frac{\Delta}{4m},\frac{\widetilde{\Delta}}{4m}\,;k,\psi\bigr)_{\ell\widetilde{\ell}}}{k}\,\Bigl(\frac{|\widetilde{\Delta}|}{\Delta}\Bigr)^{\frac{1-w}{2}}\,I_{1-w}\Bigl(\frac{\pi}{mk}\sqrt{|\widetilde{\Delta}|\Delta}\Bigr) \, , 
\end{equation}
Note that this formula in the limit of $\Delta$ going to $0$ converges and reproduces the constant terms of the Fourier expansion.
In the above, $I_\rho(x)$ is the I-Bessel function of weight $\rho$, which has the following integral representation for $x \in \mathbb{R}^*$,
\begin{equation}
\label{eq:app-Bessel}
I_\rho(x) = \frac{1}{2\pi i}\,\Bigl(\frac{x}{2}\Bigr)^\rho\,\int_{\epsilon-i\infty}^{\epsilon+i\infty}\,t^{-\rho-1}\,e^{\,t+\tfrac{x^2}{4t}}\,\mathrm{d}t \, ,
\end{equation}
and asymptotics
\begin{equation}
\label{eq:app-Bessel-asymptotic}
I_{\rho}(x) \underset{x \rightarrow \infty}{\sim} \frac{e^x}{\sqrt{2\pi x}}\Bigl(1 - \frac{\mu - 1}{8x} + \frac{(\mu - 1)(\mu - 3^2)}{2!(8x)^3} - \frac{(\mu - 1)(\mu - 3^2)(\mu - 5^2)}{3!(8x)^5} + \ldots\Bigr) \, ,
\end{equation}
with $\mu = 4\rho^2$. In \eqref{eq:app-Rad}, $Kl\bigl(\tfrac{\Delta}{4m},\tfrac{\widetilde{\Delta}}{4m}\,;k,\psi)_{\ell\widetilde{\ell}}\;$ is the generalized Kloosterman sum
\begin{equation}
Kl(\mu,\nu\,;k,\psi)_{\ell\widetilde{\ell}} \, := \sum_{\substack{0 \leq h< k \\ (h,k) = 1}}\,e^{2\pi i\,\bigl(-\tfrac{h}{k}\mu+\tfrac{h'}{k}\nu \bigr)}\;\psi(\gamma)_{\ell\widetilde{\ell}} \, ,
\end{equation}
with $\gamma = \begin{pmatrix} h' & -\tfrac{hh'+1}{k} \\ k & -h \end{pmatrix} \in SL(2,\mathbb{Z})$ and $hh' \equiv -1$ (mod $k$).


\subsection{Standard and special Jacobi forms}
\label{sec:app-special-Jacobi}

The standard Jacobi theta functions are defined as
\begin{align}
\label{eq:app-Jac-theta-1}
\vartheta_1(\tau,z) :=&\, \sum_{n\in\mathbb{Z}}\,(-1)^n\,q^{\frac12(n-\frac12)^2}\,y^{n-\frac12} \, , \\
\label{eq:app-Jac-theta-2}
\vartheta_2(\tau,z) :=&\, \sum_{n\in\mathbb{Z}}\,q^{\frac12(n-\frac12)^2}\,y^{n-\frac12} \, , \\
\label{eq:app-Jac-theta-3}
\vartheta_3(\tau,z) :=&\, \sum_{n\in\mathbb{Z}}\,q^{\frac{n^2}{2}}\,y^{n} \, , \\
\label{eq:app-Jac-theta-4}
\vartheta_4(\tau,z) :=&\, \sum_{n\in\mathbb{Z}}\,(-1)^n\,q^{\frac{n^2}{2}}\,y^{n} \, .
\end{align}
We also recall the definition of the Dedekind eta function,
\begin{equation}
\label{eq:app-eta}
\eta(\tau) := q^{\frac{1}{24}} \prod_{n\geq 1} (1-q^n) \, ,
\end{equation}
which is a modular form of weight 1/2 under $SL(2,\mathbb{Z})$ and has a multiplier system
\begin{equation}
\label{eq:app-eta-mult}
\rho_\eta(\gamma) := \exp\Bigl[i\pi \sum_{\mu\,(\mathrm{mod}\,k)}\,\Bigl(\Bigl(\frac{\mu}{k}\Bigr)\Bigr)\Bigl(\Bigl(\frac{h\mu}{k}\Bigr)\Bigr)\Bigr] \, ,
\end{equation}
with $\gamma = \begin{pmatrix} h' & -\frac{hh'+1}{k} \\ k & -h \end{pmatrix} \in SL(2,\mathbb{Z})$, $hh' \equiv -1$ (mod $k$), and 
\begin{equation}
((x)) := \begin{cases} x - \lfloor x \rfloor - \frac12 \;\; \mathrm{if} \;\; x \in \mathbb{R}\backslash\mathbb{Z} \, , \\ \qquad \qquad 0 \quad\; \mathrm{if} \;\; x \in \mathbb{Z} \, . \end{cases}
\end{equation}
Throughout the text, we denote as $\varphi_{w,m}(\tau,z)$ the standard Jacobi forms of weight $w$ and index $m$ (see \cite{Dabholkar:2012nd} for details). We report here the explicit expressions in terms of Jacobi theta functions and the Dedekind function
\begin{align}
\label{eq:app-phi-2.1}
\varphi_{-2,1}(\tau,z) :=&\, A(\tau,z) = \frac{\vartheta_1^2(\tau,z)}{\eta^6(\tau)} \, , \\
\label{eq:app-phi0.1}
\varphi_{0,1}(\tau,z) :=&\, B(\tau,z) = 4\Bigl(\frac{\vartheta_2^2(\tau,z)}{\vartheta_2^2(\tau)} + \frac{\vartheta_3^2(\tau,z)}{\vartheta_3^2(\tau)} + \frac{\vartheta_3^2(\tau,z)}{\vartheta_3^2(\tau)}\Bigr) \, , \\
\label{eq:app-phi10.1}
\varphi_{10,1}(\tau,z) =&\, \eta^{18}(\tau)\,\vartheta_1^2(\tau,z) \, .
\end{align}
The Jacobi forms $A$ and $B$ generate the ring of all weak Jacobi forms of even weight over the ring of modular forms \cite{EichZagier}.\\ \\
The Eisenstein series are defined as
\begin{align}
E_2(\tau) :=&\, 1 - 24\sum_{n=1}^\infty\,\frac{n\,q^n}{1-q^n} \, , \\
E_4(\tau) :=&\, 1 + 240\sum_{n=1}^\infty\,\frac{n^3\,q^n}{1-q^n} \, , \\
E_6(\tau) :=&\, 1 - 504\sum_{n=1}^\infty\,\frac{n^5\,q^n}{1-q^n} \, .
\end{align}
\\
\noindent We also introduce two special Jacobi forms which are needed in the discussion of the optimal choice of a mock Jacobi form for the first non-prime power index $m=6$ in section \ref{sec:mixed-Rad}.\\
The mock Jacobi form of weight 2 and index 6 $\mathcal{F}_6(\tau,z)$ is defined via
\begin{align}
\label{eq:app-F6}
&\mathcal{F}_6(\tau,z) :=\, \, \eta(\tau)\,h^{(6)}(\tau)\,\frac{\vartheta_{1}(\tau,4z)}{\vartheta_1(\tau,2z)} \, ,\\
&h^{(6)}(\tau) =\, \frac{12\,F_2^{(6)}(\tau) - E_2(\tau)}{\eta(\tau)} \, , \\
&F_2^{(6)}(\tau) =\, - \sum_{r>s>0} \chi_{12}(r^2-s^2)\,s\,q^{rs/6} \, . 
\end{align}
where $\chi_{12}(n)$ denotes the Kronecker symbol 
\begin{equation}
\chi_{12}(n) = \biggr(\frac{12}{n}\biggl) = \begin{cases} +1 \;\; \mathrm{if} \;\; n \equiv \pm1 \, (\mathrm{mod}\,12) \\ -1 \;\; \mathrm{if} \;\; n \equiv \pm5 \,(\mathrm{mod}\,12) \\ 0 \quad\; \mathrm{if} \;\; (n,12) = 1 \, . \end{cases}
\end{equation}
The mock Jacobi form of weight 2 and index 6 $\mathcal{K}_6(\tau,z)$ is
\begin{equation}
\mathcal{K}_6(\tau,z) := \frac{E_4 A B^5 - 5\,E_6 A^2 B^4 + 10\,E_4^2 A^3 B^3 - 10\,E_4 E_6 A^4 B^2 + (5\,E_4^3 - \frac14D)A^5 B - E_4^2 E_6 A^6}{12^5} \, , 
\end{equation}
with $D:=2^{11}\,3^3\,\eta^{24}(\tau)$.

\subsection{Hecke-like operators}
We define three Hecke-like operators \cite{Dabholkar:2012nd} which are needed to obtain the polar coefficients of the mixed mock modular forms appearing in section \ref{sec:mixed-Rad}.\\ \\
The first is an operator which sends a (mock) Jacobi form $\varphi(\tau,z)$ to $\varphi(\tau,sz)$,
\begin{equation}
U_s : \; \sum_{n,\ell}\,c(n,\ell)\,q^n\,y^\ell \;\; \mapsto \;\; \sum_{n,\ell}\,c(n,\ell)\,q^n\,y^{s\ell} \, ,
\end{equation}
or, in terms of its action on the Fourier coefficients of $\varphi(\tau,z)$,
\begin{equation}
c(\varphi|U_s\,;\,n,\ell) = c(\varphi\,;\,n,\ell/s) \, ,
\end{equation}
with the convention that $c(n,\ell/s) = 0$ if $s \not| \;\; \ell$.\\ \\
The second operator sends a (mock) Jacobi form of weight $w$ and index $m$ to one of weight $w$ and index $tm$ and is denoted $V_{w,t}$, and its action on the Fourier coefficients is
\begin{equation}
c(\varphi|V_{w,t}\,;\,n,\ell) = \sum_{d|(n,\ell,t)}d^{w-1}\,c\Bigl(\varphi\,;\,\frac{nt}{d^2},\frac{\ell}{d}\Bigr) \, .
\end{equation}
Finally, we define a combination of these two operators, which also sends a (mock) Jacobi form of weight $w$ and index $m$ to one of weight $w$ and index $tm$ and is given by
\begin{equation}
\label{eq:app-Hecke}
\mathcal{V}_{w,t}^{(m)} = \sum_{\substack{s^2 | t \\ (s,m) = 1}} \mu(s)\,V_{w,t/s^2}\,U_s \, ,
\end{equation}
where $\mu(s)$ is related to the M\"{o}bius function, $\mu(s) = s\,\mu_M(s)$. We have in particular the values $\mu(1) = 1$ and $\mu(2) = -2$. 

\section{Asymptotic and convergence of the circle method}
\label{sec:app-asymptotic}

To evaluate the contribution to the first term in \eqref{eq:alpha-N}, $\Sigma_1$, we split the negative and positive powers of $q$ in the expansion of $f_{m,j}$. We denote the contribution of the former by $\Sigma_1^*$ and write
\begin{equation}
\Sigma_1 = \Sigma_1^* + \sum_{\substack{0 \leq h < k \leq N \\ (h,k) = 1}} e^{-2\pi i\frac{h}{k}\frac{\Delta}{4m} + \frac{h'}{k}}\,\psi(\gamma)_{\ell j}\,\sum_{n_+ > 0} \alpha_m(n_+,j) \int_{-\vartheta_{h,k}'}^{\vartheta_{h,k}''} z^{21/2}\,e^{\frac{2\pi}{k}\bigl(z\frac{\Delta}{4m} - \frac{1}{z}\frac{\Delta^+}{4m}\bigr)} \mathrm{d}\phi \, ,
\end{equation}
where a sum over $j\in\mathbb{Z}/2m\mathbb{Z}$ is implied, $\Delta = 4mn-\ell^2$ and $\Delta^+ = 4mn_+ - j^2$. From the theory of Farey fractions, it is known that
\begin{equation}
\frac{1}{k+k_j} \leq \frac{1}{N+1} \, , \quad j \in \{1,2\} \, ,
\end{equation}
where $h_1/k_1$ is the Farey fraction antecedent $h/k$ and $h_2/k_2$ is the consecutive one. Therefore,
\begin{equation}
\vartheta'_{h,k},\,\vartheta''_{h,k} \leq \frac{1}{kN} \, .
\end{equation}
Also, recalling that $z = \frac{k}{N^2} - \mathrm{i}\,k\phi$ and that $-\vartheta_{h,k}' \leq \phi \leq \vartheta''_{h,k}$, we have the bound
\begin{equation}
|z|^2 \leq \frac{k^2}{N^4} + \frac{1}{N^2} \, .
\end{equation}
Using these results, we obtain a bound on the integral
\begin{equation}
\left|\int_{-\vartheta_{h,k}'}^{\vartheta_{h,k}''} z^{21/2}\,e^{\frac{2\pi}{k}\bigl(z\frac{\Delta}{4m} - \frac{1}{z}\frac{\Delta^+}{4m}\bigr)} \mathrm{d}\phi\right| \leq \int_{-\vartheta_{h,k}'}^{\vartheta_{h,k}''} |z|^{21/2}\,e^{\frac{\pi}{2 m N^2}\bigl(\Delta - \Delta^+|z|^{-2}\bigr)} \mathrm{d}\phi \, .
\end{equation}
Since $\Delta^+ > 0$ for $n_+ > 0$ and $j \in \mathbb{Z}/2m\mathbb{Z}$, we have that, when $N \rightarrow \infty$, the positive powers of $q$ in the expansion of $f_{m,j}$ contribute to $\Sigma_1$ a term of order
\begin{equation}
\label{eq:asymptotics-S1}
\Sigma_1 = \Sigma_1^* + \mathcal{O}\Bigl(\;\sum_{\substack{0 \leq h < k \leq N \\ (h,k) = 1}} \frac{1}{kN}\,N^{-21/2}\Bigr) = \Sigma_1^* + \mathcal{O}\Bigl(N^{-21/2}\Bigr) \, ,
\end{equation}
where the last equality follows from $\sum_{\substack{0 \leq h < k \leq N \\ (h,k) = 1}}\bigl( \frac{1}{k}\bigr)= \sum_{{0< k \leq N}}1 =N$. In conclusion, the dominant contribution to $\Sigma_1$ when $N\rightarrow \infty$ comes from the polar terms in the Fourier expansion of $f_{m,j}$. 
This polar contribution is given by
\begin{equation}
\Sigma_1^* = \sum_{\substack{\widetilde{n} \geq \widetilde{n}_0 \\ \widetilde{\ell} \in \mathbb{Z}/2m\mathbb{Z} \\ \widetilde{\Delta} < 0}} \alpha_m(\widetilde{n},\widetilde{\ell})\sum_{\substack{0 \leq h < k \leq N \\ (h,k) = 1}}e^{2\pi i(-\frac{h}{k}\frac{\Delta}{4m} + \frac{h'}{k}\frac{\widetilde{\Delta}}{4m})}\,\psi(\gamma)_{\ell\widetilde{\ell}}\, \int_{-\vartheta_{h,k}'}^{\vartheta_{h,k}''} z^{21/2}e^{\tfrac{2\pi}{k}\bigl(z\tfrac{\Delta}{4m}-\tfrac{1}{z}\tfrac{\widetilde{\Delta}}{4m}\bigr)}\;\mathrm{d}\phi \, ,
\end{equation}
with $\widetilde{\Delta} = 4m\widetilde{n}-\widetilde{\ell}^{\,2}$ and $\widetilde{n}_0$ given in \eqref{eq:n0}. We can now write the integral in terms of a Bessel function when $N\rightarrow \infty$. To do so, one needs to first write the integral in a symmetric way:
\begin{equation}
\int_{-\vartheta_{h,k}'}^{\vartheta_{h,k}''} = \int_{-\tfrac{1}{kN}}^{\tfrac{1}{kN}} - \int_{-\tfrac{1}{kN}}^{-\vartheta_{h,k}'} - \int_{\vartheta_{h,k}''}^{\tfrac{1}{kN}} \, .
\end{equation}
The second and third term contribute an error term which vanishes in the $N \rightarrow \infty$ limit \cite{BringIntegral}, and we are left with only the first integral. Integrals of these shapes can be evaluated using the method presented in \cite{BringIntegral}. For $a >0$ and $b\in\mathbb{R}^*$, they give
\begin{equation}
\label{eq:phi-integral}
\int_{-\tfrac{1}{kN}}^{\tfrac{1}{kN}} z^r\,e^{\tfrac{2\pi}{k}(a\,z + \tfrac{b}{z})}\,\mathrm{d}\phi = \begin{cases} \frac{2\pi}{k}\Bigl(\frac{b}{\sqrt{ab}}\Bigr)^{r+1}\,I_{r+1}\Bigl(\frac{4\pi}{k}\sqrt{ab}\Bigr) + \mathcal{O}\Bigl(\frac{1}{kN^{r+1}}\Bigr)\, , \quad \; \mathrm{for} \; b > 0 \,  \\[1.5mm]
\, \mathcal{O}\Bigl(\frac{1}{kN^{r+1}}\Bigr)\, ,\qquad \qquad \qquad \qquad \qquad \qquad \qquad \mathrm{for} \; b<0 \,  \end{cases}
\end{equation}
where the I-Bessel function $I_\rho(z)$ is defined in \eqref{eq:app-Bessel}. Using this result and in the limit $N\rightarrow\infty$, \eqref{eq:asymptotics-S1} shows that for $\Delta > 0$,
\begin{equation}
\Sigma_1 = \sum_{\substack{\widetilde{n} \geq \widetilde{n}_0 \\ \widetilde{\ell} \in \mathbb{Z}/2m\mathbb{Z} \\ \widetilde{\Delta} < 0}} \alpha_m(\widetilde{n},\widetilde{\ell})\, \sum_{k=1}^\infty\,\sum_{\substack{0\leq h < k \\ (h,k) = 1}}\,e^{2\pi i(-\frac{h}{k}\frac{\Delta}{4m} + \frac{h'}{k}\frac{\widetilde{\Delta}}{4m})}\,\psi(\gamma)_{\ell\widetilde{\ell}}\;\frac{2\pi}{k}\biggl(\frac{|\widetilde{\Delta}|}{\Delta}\biggr)^{23/4}\,I_{23/2}\Biggl(\frac{\pi}{m k}\sqrt{|\widetilde{\Delta}|\Delta}\Biggr) \, .
\end{equation}

\noindent We now turn to the second term in \eqref{eq:alpha-N}, the shadow contribution $\Sigma_2$. 
Before extracting its asymptotic, we write the Eichler integral $\mathcal{I}_{m,\ell}(x)$, \eqref{eq:integral} , in a new form in terms of hyperbolic functions. 
Despite the differences with \cite{Bringmann:2010sd}, we can adopt the same procedure to prove the following identity 
\be
\int_0^{\infty}\frac{\vartheta^0_{m,j} (iw-\frac{h'}{k})}{(w+x)^{3/2}} \mathrm{d}w= \sum\limits_{\substack{g\,(2mk) \\ g\equiv j(2m)}}\,e^{-\pi i \frac{g^2 h'}{2mk}}\biggl(\frac{2 \delta_{0,g}}{\sqrt{x}} - \frac{1}{\sqrt{2m}\,\pi k^2 x}\int_{-\infty}^{+\infty}e^{-2\pi x{m}u^2}f_{k,g,m}(u)\,\mathrm{d}u\biggr) \, .
\ee
Using Mittag-Leffler theory, $f_{k,g,m}(u)$ takes the form
\be
f_{k,g,m}(u):=\begin{cases} \frac{\pi^2}{\text{sinh}^2(\frac{\pi u}{k}-\frac{\pi i g}{2mk})} & \text{if} \, g\not\equiv 0 \,(\text{mod}\, 2mk) \, , \\ \frac{\pi^2}{\text{sinh}^2(\frac{\pi u}{k})}-\frac{k^2}{u^2} & \text{if} \, g\equiv 0 \,(\text{mod}\, 2mk) \, ,
\end{cases} 
\ee
This different representation of the $\mathcal{I}_{m,\ell}(x)$ integral gives rise to two contributions: one for $g\equiv 0\,(2mk)$ and one for $g \not\equiv 0\,(2mk)$. The first one has itself two contributions, coming from the polar and non-polar terms in $\eta(\tau)^{-24}$. The non-polar terms contribute an error of the type \eqref{eq:asymptotics-S1}, while the polar term $q^{-1}$ yields
\begin{equation}
\Sigma_{2,\,g\equiv0(2mk)}^* = \sum_{\substack{0 \leq h < k \leq N \\ (h,k) = 1}} \sqrt{\frac{m}{8\pi^2}}\frac{2\sqrt{k}}{\kappa}e^{2\pi i(-\tfrac{h}{k}\tfrac{\Delta}{4m}-\tfrac{h'}{k})}\,\psi(\gamma)_{\ell 0}\,\int_{-\vartheta_{h,k}'}^{\vartheta_{h,k}''} z^{11}\,e^{\tfrac{2\pi z}{k}\tfrac{\Delta}{4m}+\tfrac{2\pi}{kz}}\,\mathrm{d}\phi \, .
\end{equation}
Once again, the $\phi$ integrals are evaluated using \eqref{eq:phi-integral}, which in the limit $N\rightarrow \infty$ gives
\begin{equation}
\Sigma_{2,\,g\equiv0(2mk)} = \frac{\sqrt{2m}}{\kappa}\,\sum_{k=1}^\infty \,\sum_{\substack{0\leq h < k \\ (h,k) = 1}}e^{2\pi i(-\frac{h}{k}\frac{\Delta}{4m} - \frac{h'}{k})}\,\,\psi(\gamma)_{\ell 0}\,\frac{1}{\sqrt{k}}\biggl(\frac{4m}{\Delta}\biggr)^{6}\,I_{12}\biggl(\frac{2\pi}{k\sqrt{m}}\sqrt{\Delta}\biggr) \, .
\end{equation}
The second piece for $g \not\equiv 0\;(2mk)$ requires an analysis similar to the one conducted in \cite{Bringmann:2010sd} above Lemma 3.2. Introducing, for $b>0$ and $g \in \mathbb{Z}$,
\begin{equation}
\mathcal{J}_{k,g,m,b}(z) = e^{\frac{2\pi b}{kz}}\,z^{23/2}\,\int_{-\sqrt{b}}^{\sqrt{b}}\,e^{-\frac{2\pi m u^2}{kz}}\,f_{k,g,m}(u)\,\mathrm{d}u \, ,
\end{equation}
one can show that, in the limit $N\rightarrow \infty$,
\begin{align}
\Sigma_{2,\,g\not\equiv0(2mk)}^* =&\, -\frac{1}{4\pi^2 \kappa}\sum_{\substack{0 \leq h < k \leq N \\ (h,k) = 1}}\frac{1}{k}\sum_{\substack{j\in\mathbb{Z}/2m\mathbb{Z} \\ g(2mk) \\ g \equiv j (2m)}}e^{2\pi i\bigl(-\tfrac{h}{k}\tfrac{\Delta}{4m}-\tfrac{h'}{k}\bigl(1+\tfrac{g^2}{4m}\bigr)\bigr)}\,\psi(\gamma)_{\ell g} \; \times \nonumber \\
&\qquad \int_{-\vartheta_{h,k}'}^{\vartheta_{h,k}''} e^{\tfrac{2\pi}{k}z\tfrac{\Delta}{4m}}\mathcal{J}_{k,g,m,1}(z)\,\mathrm{d}\phi \, .
\end{align}
We now evaluate the $\phi$ integral as usual. Using \eqref{eq:phi-integral} and in the limit $N\rightarrow \infty$, this truncates the integration range over $u$ to the region where $1-mu^2$ is positive, i.e.
\begin{align}
\Sigma_{2,\,g \not\equiv 0(2mk)} =&\, -\frac{1}{2\pi \kappa}\,\sum_{k=1}^\infty\,\sum_{\substack{j\in\mathbb{Z}/2m\mathbb{Z} \\ g(2mk) \\ g \equiv j (2m)}}\,\sum_{\substack{0\leq h < k \\ (h,k) = 1}}e^{2\pi i\bigl(-\tfrac{h}{k}\tfrac{\Delta}{4m}-\tfrac{h'}{k}\bigl(1+\tfrac{g^2}{4m}\bigr)\bigr)}\,\psi(\gamma)_{\ell g}\,\frac{1}{k^2}\biggl(\frac{4m}{\Delta}\biggr)^{25/4}\,\times \nonumber \\
&\,\int_{-1/\sqrt{m}}^{+1/\sqrt{m}}\,f_{k,g,m}(u)\,I_{25/2}\Biggl(\frac{2\pi}{k\sqrt{m}}\sqrt{\Delta(1-m u^2)}\Biggr)(1-m u^2)^{25/4}\,\mathrm{d}u \, .
\end{align}

\section{Edge-effects in supergravity}
\label{sec:app-Iu}

In this appendix, we perform a detailed analysis of the~$I_u$ integral which appeared in the computation of the quantum entropy function for 1/4-BPS black holes in $\mathcal{N}=4$ supergravity,
\begin{equation}
I_u(p,\bar{p}) := \int_{\gamma_1}\,\mathrm{d}\tau_1\,\exp\Bigl[\frac{\pi m}{\tau_2}\bigl(\tau_1 + i(p-\bar{p})\frac{\tau_2}{m} - \frac{\ell}{2m}\bigr)^2\Bigr] \, .
\end{equation}
This integral is then multiplied by a $\tau_2$-dependent prefactor and integrated over a contour $\gamma_2$ as in \eqref{eq:QEF-perfect-square}. In \cite{Murthy:2015zzy}, the contours $\gamma_1$ and $\gamma_2$ were chosen to be (see also \cite{Gomes:2015xcf})
\begin{align}
&\,\tau_1 = i\,\tau_2\,u \;\; : \;\; -1+\delta \leq u \leq 1-\delta \, , \\
&\,\tau_2 \qquad \qquad : \;\; \epsilon - i\infty < \tau_2 < \epsilon+ i\infty \, ,
\end{align}
with $\delta$ small and positive and $\epsilon$ strictly positive. On the contour~$\gamma_1$, we have:
\begin{align}
I_u(p,\bar{p}) =&\, i\tau_2\int_{-1+\delta}^{1-\delta}\,\mathrm{d}u\,\exp\Bigl[-\pi m\tau_2(u + \alpha - \frac{\ell}{2im\tau_2})^2\Bigr] \, , \\
=&\, \frac{1}{2}\sqrt{\frac{\tau_2}{m}}\Bigl[\mathrm{Erfi}\bigl(\frac{\sqrt{\pi}(\ell-2i\tau_2 m(\alpha-1+\delta)}{2\sqrt{\tau_2 m}}\bigr) - \mathrm{Erfi}\bigl(\frac{\sqrt{\pi}(\ell-2i\tau_2 m(\alpha+1-\delta)}{2\sqrt{\tau_2 m}}\bigr)\Bigr] \, , \nonumber
\end{align}
where~$\text{Erfi}(x)$ is the imaginary error function. We have also defined $\alpha := (p-\bar{p})/m$.\\ \\
Taking~$\text{Re}(\tau_2) = \epsilon$ to be very large and using the Taylor series of the imaginary error function in this regime, one obtains three results depending on the value of~$|\alpha|$. First for~$|\alpha| < 1-\delta$,
\begin{align}
\label{Iualphaless}
I^{|\alpha|< 1-\delta}_u = i\sqrt{\frac{\tau_2}{m}}\,+&\, \exp\Bigl[\pi\frac{(\ell-2i\tau_2 m(\alpha-1+\delta))^2}{4\tau_2 m}\Bigr]\Bigl(\frac{i}{2\pi m(\alpha-1+\delta)} + \mathcal{O}\bigl(\frac{1}{\epsilon}\bigr)\Bigr) \cr
-&\, \exp\Bigl[\pi\frac{(\ell-2i\tau_2 m(\alpha+1-\delta))^2}{4\tau_2 m}\Bigr]\Bigl(\frac{i}{2\pi m(\alpha+1-\delta)} + \mathcal{O}\bigl(\frac{1}{\epsilon}\bigr)\Bigr) \, .
\end{align}
Second, for~$|\alpha| = 1-\delta$:
\begin{equation}
\label{Iualphaequal}
I^{\alpha = \pm(1-\delta)}_u = \frac{i}{2}\sqrt{\frac{\tau_2}{m}} \pm \frac{\ell}{2m} - \exp\Bigl[\pi\frac{(\ell\pm4i\tau_2 m(\delta-1))^2}{4\tau_2 m}\Bigr]\Bigl(\frac{i}{4\pi m(1-\delta)} + \mathcal{O}\bigl(\frac{1}{\epsilon}\bigr)\Bigr) \, .
\end{equation}
Third, for~$|\alpha| > 1-\delta$:
\begin{align}
\label{Iualphamore}
I^{|\alpha| > 1-\delta}_u =&\, \exp\Bigl[\pi\frac{(\ell-2i\tau_2 m(\alpha-1+\delta))^2}{4\tau_2 m}\Bigr]\Bigl(\frac{i}{2\pi m(\alpha-1+\delta)} + \mathcal{O}\bigl(\frac{1}{\epsilon}\bigr)\Bigr) \cr
&\,-\exp\Bigl[\pi\frac{(\ell-2i\tau_2 m(\alpha+1-\delta))^2}{4\tau_2 m}\Bigr]\Bigl(\frac{i}{2\pi m(\alpha+1-\delta)} + \mathcal{O}\bigl(\frac{1}{\epsilon}\bigr)\Bigr) \, .
\end{align}
We can use the above expressions for~$I_u$ in~\eqref{eq:QEF-perfect-square}. The~$\tau_2$ integral is now on a contour where~$\epsilon \gg 1$, but since the only pole in the~$\tau_2$ complex plane sits at the origin, we can safely deform it back to~$\epsilon$ small and still positive. This then gives rise to I-Bessel functions of weight 23/2 using \eqref{eq:app-Bessel}, as well as what \cite{Murthy:2015zzy} called edge-effects, coming from taking the limit $\delta \rightarrow 0$. The contribution of the latter can be written as I-Bessel functions of weight 12, and it is given by
\begin{align}
\label{eq:app-QEF-Bessel12}
\widehat{W}^{I_{12}}(n,\ell,m) =&\,\displaystyle{\sum_{\substack{p \geq -1 \\ p+1\,\neq\,m}} (m-p+1)d(p)\frac{2\cos(\pi\ell)}{p+1-m}\Bigl(\frac{4m}{4mn}\Bigr)^6\,I_{12}\Bigl(\frac{2\pi}{\sqrt{m}}\sqrt{4mn}\Bigr)} \nonumber \\
&\,+\frac{1}{m}\sum_{-1 \leq p < m}\,p\,d(p)\,d(p + m)\,e^{i\pi\ell}\Bigl(\frac{|4p+4m|m}{4mn}\Bigr)^{6}\,I_{12}\Bigl(\frac{2\pi}{\sqrt{m}}\sqrt{|p+m|4mn}\Bigr) \nonumber \\
&\,\displaystyle{+\frac{4\pi\ell}{m}\,\sin(\pi\ell)\,d(m-1)\Bigl(\frac{4m}{4mn-\ell^2}\Bigr)^6\,I_{12}\Bigl(\frac{2\pi}{\sqrt{m}}\sqrt{4mn-\ell^2}\Bigr)} \, ,
\end{align}
where $d(p)$ is the $p^\mathrm{th}$ Fourier coefficient of $\eta(\tau)^{-24}$. We see from this expression that the edge-effects actually give rise to an infinite series of I-Bessel functions of weight 12. One could hope that this contribution can be rewritten in the form of a single I-Bessel function of weight 12 along with an integral of an I-Bessel function of weight 25/2 as the one appearing in \eqref{eq:mixed-mock-coeffs}. This would strengthen the match between the supergravity calculation and the Fourier coefficients of mixed mock modular forms. However at present, it is not clear if such a form can be achieved from \eqref{eq:app-QEF-Bessel12}.

\end{appendix}
\newpage

\end{document}